\documentclass[lettersize,journal]{IEEEtran}
\usepackage{amsmath,amsfonts}

\usepackage[linesnumbered, ruled]{algorithm2e}
\SetKwRepeat{Do}{do}{while}%
\usepackage{cite}
\usepackage{array}

\usepackage{tabularx}
\usepackage{booktabs}

\usepackage{CJKutf8} 
\usepackage{bbding}
\usepackage[caption=false,font=normalsize,labelfont=sf,textfont=sf]{subfig}
\usepackage{multirow}
\usepackage{makecell}
\usepackage{textcomp}
\usepackage{stfloats}
\usepackage{url}
\usepackage{verbatim}
\usepackage{graphicx}
\usepackage[T1]{fontenc} 
\usepackage{amssymb} 
\usepackage[backref=False]{hyperref}
\def\BibTeX{{\rm B\kern-.05em{\sc i\kern-.025em b}\kern-.08em
    T\kern-.1667em\lower.7ex\hbox{E}\kern-.125emX}}
\usepackage{balance}
\newcommand{\Rmnum}[1]{\uppercase\expandafter{\romannumeral #1}}  
\usepackage{tikz,xcolor,hyperref}

\usepackage{amsmath}
\usepackage{algpseudocode}
\usepackage{subcaption}
\usepackage{graphicx}
\usepackage{amssymb}

\usepackage{tcolorbox}   
\tcbset{
	llm-explain/.style={
		colback=gray!10,    
		colframe=blue!30,   
		fonttitle=\bfseries,
		title={LLM生成的解释}, 
		left=1em, right=1em, top=0.5em, bottom=0.5em 
	}
}

\definecolor{lime}{HTML}{A6CE39}
\DeclareRobustCommand{\orcidicon}{
	\begin{tikzpicture}
		\draw[lime, fill=lime] (0,0)
		circle[radius=0.16]
		node[white]{{\fontfamily{qag}\selectfont \tiny \.{I}D}}; 
	\end{tikzpicture}
	\hspace{-2mm}
}
\foreach \x in {A, ..., Z}{%
	\expandafter\xdef\csname orcid\x\endcsname{\noexpand\href{https://orcid.org/\csname orcidauthor\x\endcsname}{\noexpand\orcidicon}}
}

\begin{document}
\title{NLP-Based Review for Toxic Comment Detection Tailored to the Chinese Cyberspace}

\author{Ruixing~Ren\hspace{-1.5mm}\orcidA{},~\IEEEmembership{Graduate~Student~Member,~IEEE}, Junhui~Zhao\hspace{-1.5mm}\orcidC{},~\IEEEmembership{Senior~Member,~IEEE}, Xiaoke~Sun\hspace{-1.5mm}\orcidD{}, and Qiuping~Li\hspace{-1.5mm}\orcidE{}

\thanks{Corresponding author: Junhui Zhao. 
Ruixing Ren, Junhui Zhao are with the School of Electronic and Information Engineering, Beijing Jiaotong University, Beijing 100044, China. (e-mail: renruixing0604@163.com)
	
Xiaoke Sun and Qiuping Li are with the National Computer Network Emergency Response Technical Team/Coordination Center of China (CNCERT/CC), Beijing 100029, China.}
}
\maketitle

\begin{abstract}
  With the in-depth integration of mobile Internet and widespread adoption of social platforms, user-generated content in the Chinese cyberspace has witnessed explosive growth. Among this content, the proliferation of toxic comments poses severe challenges to individual mental health, community atmosphere and social trust. Owing to the strong context dependence, cultural specificity and rapid evolution of Chinese cyber language, toxic expressions are often conveyed through complex forms such as homophones and metaphors, imposing notable limitations on traditional detection methods. To address this issue, this review focuses on the core topic of natural language processing based toxic comment detection in the Chinese cyberspace, systematically collating and critically analyzing the research progress and key challenges in this field. This review first defines the connotation and characteristics of Chinese toxic comments, and analyzes the platform ecology and transmission mechanisms they rely on. It then comprehensively reviews the construction methods and limitations of existing public datasets, and proposes a novel fine-grained and scalable framework for toxic comment definition and classification, along with corresponding data annotation and quality assessment strategies. We systematically summarize the evolutionary path of detection models from traditional methods to deep learning, with special emphasis on the importance of interpretability in model design. Finally, we thoroughly discuss the open challenges faced by current research and provide forward-looking suggestions for future research directions.
\end{abstract}


\section{Introduction}
Mobile communication technology, especially the generational leap from 1G to 5G, has fundamentally reshaped how human societies connect and exchange information \cite{ZhaoHetNets,RenIOTJ}. At the core of this technological revolution is the liberation of communication from fixed locations, along with the continuous enhancement of bandwidth, speed and reliability \cite{Ren6G}, which has fostered the new paradigm of mobile Internet. This has led to a profound evolution of the Internet: from an information network in the desktop era to a people- and social-centric living platform and ubiquitous environment in the mobile age.

This evolution means not only a shift in access devices, but also a reshaping of the Internet ecosystem. Social media, short-video platforms and instant messaging applications have become primary channels for information dissemination and public interaction, with an explosive growth in user-generated content and communication turning real-time, fragmented and highly contextualized \cite{Smartphone}. While unleashing enormous creativity, this ecological transformation also poses a series of complex governance challenges \cite{Zaitsev}. Platforms must address unprecedented issues such as information authenticity (e.g., rumors \cite{FakeNews}), interaction health (e.g., toxic comments \cite{TOXICN}) and information distribution mechanisms (e.g., algorithmic recommendation \cite{Recommendation}) to uphold order, security and trust in cyberspace.

Against this backdrop, the detection and governance of toxic comments, including insults, hate speech and personal attacks, have emerged as one of the key technical priorities in global Internet governance, owing to their direct harm to individual mental health, community atmosphere and social consensus. This has spurred extensive research in multilingual contexts. Currently, research based on English \cite{English1,English2,English3} is the most in-depth and comprehensive, with corresponding progress also made in languages such as French \cite{French}, Arabic \cite{Arabic} and Turkish \cite{Turkish}.

In recent years, with the expansion of the user base and social influence of Chinese-language Internet, research on the detection of Chinese toxic comments has attracted growing attention. We review the key advances in this field chronologically in Table \ref{tab1} \footnote{Constrained by the table length, only several representative works on sarcasm are presented. Meanwhile, since this review focuses on toxic comments, many sarcastic comments are non-toxic, such as self-mockery.}, covering research scopes, dataset sources, scales, balance and openness, as well as the design of toxic classification levels. Nevertheless, it must be acknowledged that the detection of Chinese toxic comments faces formidable challenges stemming from the uniqueness of the language and online culture. The difficulties go far beyond superficial grammatical structure or keyword recognition. The in-depth challenges are rooted in the unique, dynamic and highly contextualized subcultural ecosystem of the Chinese Internet \cite{Xiao2024ChineseOL}. This ecosystem has fostered a wealth of indirect, symbolic and ever-evolving modes of expression. Users commonly resort to homophones \cite{HED-COLD}, irony \cite{CSCET}, metaphor \cite{MCIHD} and cultural allusions \cite{Diangu} to veil aggressive intentions, leading to a stark disconnect between literal meanings and actual sentiments. Meanwhile, the deep integration of memes \cite{multimodality}, Internet catchphrases, dialectal variants and community-specific jargon often conceals toxic content beneath a veneer of humor or in-group culture, requiring detection systems to possess cross-modal comprehension and in-depth cultural background knowledge.

Furthermore, the rapid generation and dissemination characteristics of Chinese online language lead to the constant emergence of new taboo words, variant words and aggressive metaphors, subjecting detection models to severe generalization and timeliness pressures \cite{STATE_ToxiCN}. More importantly, the meaning of such expressions is highly dependent on specific platforms, topic threads and immediate contexts \cite{LI2020101453}; the same sentence may carry drastically different emotions and intentions across scenarios such as Weibo, Douyin and Zhihu. This deep reliance on complex contexts and dynamic cultures constitutes a distinctive particularity that differentiates Chinese toxic comment detection from that in other languages \cite{10890580}.

Thus, to address the aforementioned unique challenges and systematically sort out the development trajectory of this field, this review aims to comprehensively summarize and evaluate the research on toxic comment detection for the Chinese cyberspace against the backdrop of the rapid advancement of artificial intelligence (AI) and natural language processing (NLP) technologies. In recent years, AI technologies, represented by deep learning (DL), have markedly enhanced machines' ability to understand and generate complex linguistic patterns, enabling automated content moderation and governance \cite{RenUAV,RenIoV,RenITS}. Among these, NLP, as a key technology to enable machines to comprehend human language, has achieved breakthroughs in tasks such as sentiment analysis, intent recognition and harmful information detection, providing a core tool for cyberspace governance.

\begin{figure*}[t]
	\centerline{\includegraphics[width=7.5in,keepaspectratio]{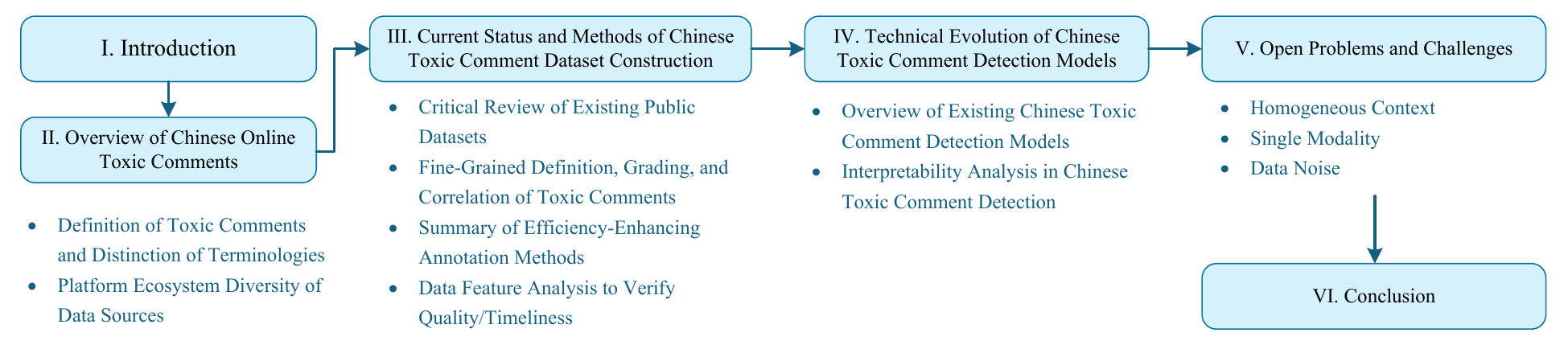}}
	\caption{Structure of the Review and Topics Discussed in Each Section.}
	\label{fig0}
\end{figure*}

Amid this technological wave, toxic comment detection has emerged as a crucial and challenging application area in the NLP field. However, existing studies mostly focus on languages such as English, and systematic summaries targeting Chinese, a large and unique context, remain insufficient. To the best of our knowledge, this review is the second systematic literature survey dedicated to Chinese online toxic comment detection following the work of Xiao et al \cite{Xiao2024ChineseOL}. Compared with previous work, we strive to provide a more comprehensive, specific and in-depth discussion in terms of coverage breadth, analysis depth and forward-looking perspective on cutting-edge challenges. We organize this review following the logic of problem definition, data foundation, technical methods, open challenges, and the specific contributions and structure of this review are as follows:
\begin{itemize}
	\item Section \ref{2} provides an overview of Chinese online toxic comments, clarifies their specific definitions and lexical boundaries in the Chinese context, analyzes the diverse platform ecosystems where such comments emerge and spread, and lays the conceptual and contextual foundation for subsequent discussions.
	\item Section \ref{3} focuses on the status quo and methods of dataset construction for Chinese toxic comments. It not only critically reviews existing public datasets, but also proposes a novel fine-grained definition and classification framework for toxic comments, along with efficient data annotation methods based on this framework and post-annotation data quality assessment approaches, thereby providing data-level insights for model development.
	\item Section \ref{4} systematically reviews the technical evolution of Chinese toxic comment detection models, outlines the development path of mainstream detection methods, and highlights the crucial direction of model interpretability, so as to improve the reliability and transparency of detection systems.
	\item Section \ref{5} further analyzes the core open issues and structural challenges faced by current research, including the models' over-reliance on homogeneous contexts, insufficient utilization of multimodal information, and the impact of data noise on system performance, thus identifying the key bottlenecks to be addressed in future research. Section \ref{6} concludes the review.
\end{itemize}

The detailed content structure of each section is shown in Fig. \ref{fig0}. This review aims to provide a clear roadmap and solid literature foundation for constructing a toxic comment detection system better adapted to the complexity of the Chinese cyberspace, through systematic collation and critical analysis of existing studies.

\begin{table*}[htbp]
	\centering
	\caption{Summary of Chinese Toxic Detection Datasets}
	\label{tab1}
	\renewcommand{\arraystretch}{1.2}
	\begin{tabular}{|m{1.9cm}|m{1.3cm}|m{7.5cm}|m{0.8cm}|m{0.8cm}|m{1cm}|m{0.6cm}|}
		\hline
		\textbf{Dataset Name} & \textbf{Source} & \textbf{Research Scope} & \textbf{Label} & \textbf{Scale} & \textbf{Balance} & \textbf{Avail.} \\
		\hline
		NTU Irony \cite{NTU_Irony} & Plurk & Construction of an Ironic Corpus; Mining Ironic Linguistic Patterns and Structural Elements & 2 & 1,005 & 100\% & \Checkmark \\
		\hline
		RPCT \cite{RPCT} & Twitter, PTT & Constructing a Rule-Based System for Profanity Detection and Rewriting & 2 & 9,557 & >50\% & \Checkmark \\
		\hline
		Gong et al. \cite{Gong2020} & Guanchazhe & Sarcasm Data Construction and Enhancing Data Effectiveness by Integrating News Context & 2 & 4,972 91,782 & 1:1 2.71\% & \XSolidBrush  \\
		\hline
		Li et al. \cite{Li2020} & Weibo &  Extracting Ironic Constructions and Processes Based on Reversal Theory for Irony Detection& 2 & - & - & -  \\
		\hline
		COLA \cite{COLA} & YouTube, Weibo & Construction of a Fine-Grained Dataset for Offensive Language and Network Design to Improve Classification Performance & 4 & 18,707 & $\approx$54\% & \XSolidBrush  \\
		\hline
		TOCP \cite{TOCP} & Twitter, PTT & Improving the Quality of Profanity Detection and Handling & 2 & 16,450 & $\approx$86.8\% & \Checkmark  \\
		\hline
		Zhang et al. \cite{Zhang2022} & Internet Forum & Proposing a Hybrid Model of Character + Word-Level Feature Fusion for Toxic Comment Detection & 2 & 43,789 & $\approx$43.7\% & \XSolidBrush \\
		\hline
		CDIAL-Bias \cite{COIAL-Bias} & Zhihu & Social Bias Detection in Dialogue Systems and Proposing a Multi-Dimensional Analysis Framework & 4  & 13,394 15,013 & 18.9\% 25.9\% & \Checkmark \\
		\hline
		SWSR \cite{SWSR} & Weibo & Online Sexism Detection and Supporting Multi-Granularity Classification and Cross-Domain Research & 2  & 8,969 & $\approx$34.5\% & \Checkmark \\
		\hline
		COLD \cite{COLD} & Zhihu, Weibo & Construction of Offensive Language Detection Benchmark and Improving the Accuracy of Detection Model & 2 & 37,480 & $\approx$48.1 & \Checkmark  \\
		\hline
		CSarc-Context \cite{CSarc-Context} & Social Network & Proposing a Retrospective Reader Model Fusing Context for Sarcasm Detection  & 2 & Large & - & - \\
		\hline
		Zhu et al. \cite{Zhu2022} & Weibo etc. & Construction of an Open Sarcasm Corpus and Achieving High Data Density and Quality Control & 2 & 2,000 & 1:1 & \Checkmark \\
		\hline
		TOXICN \cite{TOXICN} & Zhihu, Tieba & Fine-Grained Toxic Language Detection and Proposing a Hierarchical Classification Framework & 2 & 12,011 & $\approx$54.8\% & \Checkmark \\
		\hline
		Li et al. \cite{Li2023} & COLD-Based & Proposing a Three-Level Feature Fusion Model of Character-Word-Sentence for offensive Text Detection & 2 & 25,820 & $\approx$48.1\% & - \\
		\hline
		AugCOLD \cite{AugCOLD} & Prompt Gen. etc. & Improving Model Generalization by Data Augmentation and Knowledge Distillation for Detection & 2 & 1,090k & $\approx$42\% & \Checkmark \\
		\hline
		Hou et al. \cite{Hou2024} & COLD-Based & Enhancing Model Capture of Deep Semantics and Local Key Information for Offensive Language Detection & 2 & 37,480 & $\approx$48.1\% & - \\
		\hline
		ToxiCloakCN \cite{ToxiCloakCN} & ToxiCN-Based & Robustness Evaluation and Shortcomings Analyzing of Offensive Language Detection Model & 2 & 4,582 & $\approx$50.1\% & \Checkmark \\
		\hline
		Cao et al. \cite{Cao2024} & COLD Extension &  Improving Offensive Language Detection Performance by Fusing Semantic and Thematic Features & 2 & 37,480 10,717 & $\approx$48.1\% & \Checkmark \\
		\hline
		MCIHD \cite{MCIHD} & Social Platform & Adapting to Low-Resource Scenarios by Domain-Enhanced Prompt Learning for Implicit Hate Speech Detection & 3 & 20,000 & $\approx$36.9\% & \XSolidBrush \\
		\hline
		Observers Corpus \cite{Observers_Corpus} & Guanchazhe & Optimizing Detection Effect by Fusing Title and Comment Information for Sarcasm Recognition & 2 & 4,972 & 1:1 & \Checkmark \\
		\hline
		Ren et al. \cite{Ren2024} & Public Dataset & Alleviating Model Overfitting Through Loss Function Optimization  for Sarcasm Detection Model & 2 & 10,380 & $\approx$41.7\% & \Checkmark \\
		\hline
		SCCD \cite{SCCD} & Weibo & Session-Level Cyberbullying Detection; Analyzing the Impact of Session Context on Detection & 2 & 677 38,999 & $\approx$52.3\% 25.1\% & \Checkmark \\
		\hline
		MCSh \cite{MCSh} & Talk Show & Multimodal Sarcasm Detection; Improving Recognition Capability by Fusing Audio and Text Features & 2 & 6,124 & 20\% & \Checkmark \\
		\hline
		CSCET \cite{CSCET} & Weibo & Enhancing Feature Distinguishability by Incorporating Event Context for Sarcasm Detection & 2 & 30,327 & $\approx$64.3\% & \Checkmark \\
		\hline
		CTSD \cite{CTSD} & Douyin & Proposing a Hybrid Deep Learning Model Adapted to the Characteristics of Online Language for Sarcasm Detection & 2 & 28,630 & $\approx$27.9\% & \XSolidBrush \\
		\hline
		Zhang et al. \cite{Zhang2025} & Bilibili &  Fusing Target Comments and Multi-Dimensional Contextual Information for Sarcastic Comment Detection & 2 & 79,045 5,628 & 3.56\% 1:1 & \Checkmark \\
		\hline
		MCSD \cite{MCSD} & Talk Show & Cross-Lingual Multimodal Sarcasm Detection; Verifying the Detection Advantages of Multimodal Fusion & 2 & 2,725 & 1:1 & \Checkmark \\
		\hline
		Offensive\_Data \cite{Offensive_Data} & COLD ToxiCN & Improving Training Efficiency and Generalization by Parallel Multi-Task Learning for Offensive Language Detection & 2 & 47,490 & $\approx$49.3\% & \XSolidBrush \\
		\hline
		CangjieToxi  \cite{CangjieToxi} & Douyin & Offensive Language Anti-Evasion Detection; Analyzing Model Vulnerability to Structural Perturbations & 2 & 28,080 & $\approx$49.9\% & \Checkmark \\
		\hline
		ChineseHarm Bench  \cite{ChineseHarm-Bench} & Violation Record & Multi-Category Harmful Content Detection; Reducing Model Deployment Costs Through Knowledge Enhancement & 6 & 6,000 & $\approx$83.3\% & \Checkmark \\
		\hline
		STATE ToxiCN  \cite{STATE_ToxiCN} & ToxiCN-Based & Fine-Grained Hate Speech Detection; Realizing Span-Level Target and Attribute Extraction & 4 & 8,029 & $\approx$63.6\% & \Checkmark \\
		\hline
		TE-Dataset \cite{TE-Dataset} & Bilibili & Enhancing Model Stability in Cross-Topic and Noisy Scenarios for Toxic Euphemism Detection & 2 & 18,971 & $\approx$55.7\% & \Checkmark \\
		\hline
		HED-COLD \cite{HED-COLD} & COLD-Based & Enhancing Offensive Detection Models' Recognition of Homophonic Perturbation & 2 & 10,000 & $\approx$70\% & \Checkmark \\
		\hline
	\end{tabular}
\end{table*}

\section{Overview of Chinese Online Toxic Comments}\label{2}

\subsection{Definition of Toxic Comments and Distinction of Terminologies}\label{2.1}
This review focuses on "comment" in Chinese cyberspace, specifically referring to short user-generated texts attached to news, videos, social media posts, or blog entries. The selection of this core research object aims to clearly distinguish between broader concept of "language" and the context-specific term "speech".

From the perspectives of linguistics and communication studies, "language" is a broad system that encompasses abstract rules such as vocabulary, grammar, and pragmatics, as well as all speech acts ranging from formal documents to private conversations. In contrast, "speech" is inherently associated with the fundamental right to freedom of expression. Within legal and political philosophy discourses, it embodies connotations related to public expression, the marketplace of ideas, and the boundaries of power, typically referring to more public, purposeful, and complete statements of opinion such as public speeches, political declarations, and editorial articles.

In comparison, the "comment" focused on in this review exhibit uniqueness in the context of computer-mediated communication. (1) Dependence and interactivity: deeply embedded in the context of original content to form conversational chains, serving as a direct manifestation of community interaction \cite{Waterschoot}; (2) Immediacy and fragmentation: generated without deliberate deliberation, mostly short and casual texts that reflect users’ immediate emotions \cite{Madhyastha}; (3) Extensive participation: acting as the most direct public expression tool for the vast majority of internet users, and forming the basic unit of the online conversational ecosystem \cite{Thomas}. These characteristics render comment sections a primary breeding ground for toxic content: anonymity lowers the threshold for aggression, lack of context exacerbates misunderstandings, and dynamic interactions tend to escalate conflicts. Therefore, focusing on online comments rather than the generalized online language or speech can make research objectives more precise and hold greater practical value for purifying the Chinese online communication environment.

Prior to elaborating on the discussion, it is necessary to clarify the core operational concept of this study. A plethora of terms describing negative online content exist in the existing literature and are frequently conflated, such as toxic/toxicity, offensive/offense, hate, harmful, profanity, bias, cyberbullying, abusive, sexism, abuse, aggressiveness, misogyny, and sarcasm/irony. These terms exhibit subtle differences in aggression intensity, target orientation, and scope of coverage, forming a complex conceptual spectrum \cite{JAHAN2023126232}. To establish a unified and clear analytical framework, precise discrimination of these terms is required.
\begin{itemize}
	\item \textbf{Offensive}: Refers to content that elicits disgust or emotional discomfort in others, with relatively subjective judgment criteria. A comment may be offensive yet not necessarily toxic (e.g., severe yet rational criticism), and its scope is narrower than that of toxicity.
	\item \textbf{Hate}: Specifically refers to aggressive comments based on inherent characteristics of individuals or groups such as race, religion, and gender. It is a typical and severe subclass of toxic content, yet fails to cover ordinary insults or cyberbullying without specific biased targets.
	\item \textbf{Harmful}: A consequence-oriented concept referring to content that may cause actual harm. Its scope is overly broad, encompassing all toxic comments as well as disinformation, terrorist propaganda, and the like. Some content may be harmful but not toxic (e.g., rumors), and this concept highly overlaps with legal and political domains, which is beyond the scope of this review.
	\item \textbf{Profanity}: Toxic content directly manifested through offensive words such as obscenities and vulgarities, serving as the most direct and explicit linguistic marker in toxic comments. Even veiled attacks without obscenities qualify as toxic, but profane content, due to the extreme offensiveness of its wording, inherently significantly undermines the conversational environment.
	\item \textbf{Bias}: Refers to aggressive or derogatory comments made based on inherent prejudices and pre-judgments towards specific groups or individuals. Unlike internal biased attitudes, this concept specifically denotes the externalization of prejudice into harmful verbal behaviors, and it serves as a core component of systematic toxic content such as hate comments.
	\item \textbf{Cyberbullying}: Malicious attacks targeted at specific individuals, usually characterized by repetitiveness. Its core lies in behavioral patterns, and it cannot cover toxic comments without specific targets or extensive group-directed hate comments.
	\item \textbf{Abusive}: Focuses on direct and explicit personal attacks and insults, serving as a major manifestation of toxicity. However, it fails to cover implicit and indirect forms of toxicity.
	\item \textbf{Sexism \& Misogyny}: Specific subcategories of hate speech that cannot cover aggressive behaviors targeting other characteristics such as race and region.
	\item \textbf{Aggressive}: Describes a confrontational and provocative tone, and it is more of an attribute or manifestation of toxicity rather than a complete classificatory label.
	\item \textbf{Sarcasm/Irony}: Its core lies in semantic contrast (where the literal meaning contradicts the true intention), and it is not inherently harmful (e.g., a friend’s tease: \begin{CJK}{UTF8}{gbsn} 你这电影选得可真好，我睡了最香的一觉\end{CJK} [Nice choice on the movie. I had the best nap of my life.]). However, when sarcasm aims at belittlement, discrimination, or humiliation and targets specific individuals or groups, it falls into the category of toxic comments.
\end{itemize}
As revealed by the above analysis, all these terms have limitations: they are either overly broad (e.g., Harmful), excessively narrow and specific (e.g., Hate Speech, Cyberbullying), focused solely on superficial characteristics (e.g., Profanity), or centered on internal attitudes (e.g., Bias). None of them can meet the needs of this review for the precise definition of core concepts.

Therefore, this review identifies "toxicity" as the top-level overarching concept. This concept originates from the pioneering work of Google Jigsaw, which defines toxic comments as "rude, disrespectful, or unreasonable comments that are likely to drive people away from a conversation" \cite{Jigsaw, Toxicity_define}. The core advantage of this definition lies in the shift of perspective: it moves the analytical focus from the specific content of comments ("what is said") to their expression methods and social impacts ("how it is said" and "what effects it produces"). This shift enables it to transcend narrow definitions based on specific themes (e.g., politics, gender) and provides a neutral and solid conceptual foundation for constructing a universal and scalable detection model.

\begin{figure*}[t]
	\centerline{\includegraphics[width=7.0in,keepaspectratio]{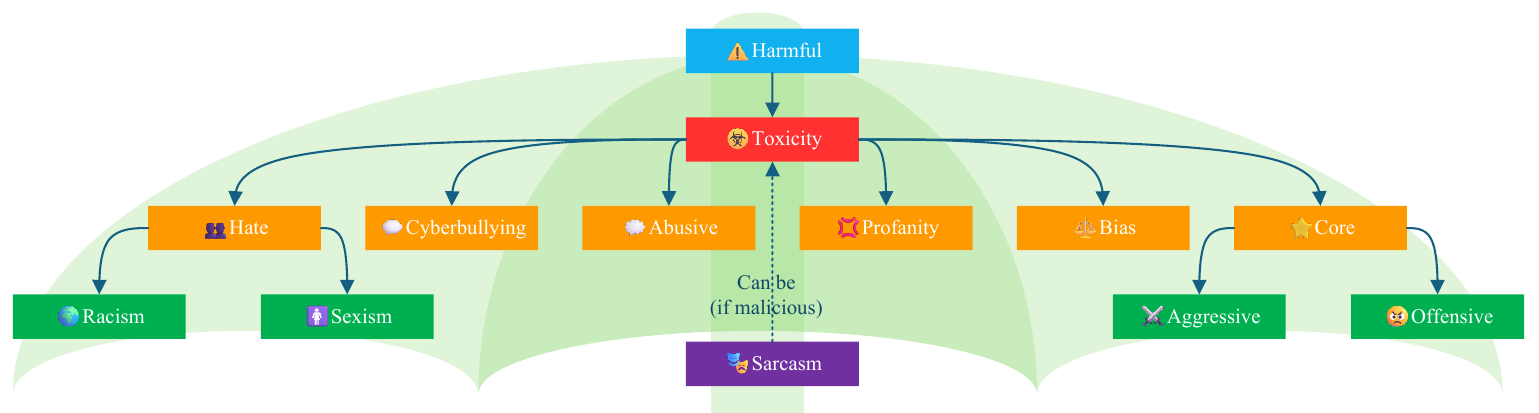}}
	\caption{Conceptual Hierarchy of Toxic Comment.}
	\label{fig1}
\end{figure*}

The systematicity and hierarchy of this conceptual framework can be clearly visualized through Fig. \ref{fig1}. As shown in the figure, Toxicity is not a flat label but a hierarchical system with inherent logic. On the one hand, as an umbrella term, it encompasses a series of specific behaviors such as Profanity, Abusive, Cyberbullying, and Hate. On the other hand, it reveals that these behaviors share core attributes like Aggressiveness and Offensiveness, and Toxicity itself is subordinate to the broader category of Harmful. As an independent linguistic phenomenon, Sarcasm overlaps with the category of toxic comments only when it serves the intention of attacking, demeaning, or disrupting conversations.

To summarize, relying on its neutral perspective and clear structure, the concept of toxicity precisely defines the operational scope of this review, transforming complex online negative behaviors into structured problems that can be understood and processed by computational models. This not only establishes clear targets for the subsequent development of detection technologies but also paves the way for systematic evaluation and governance.

\subsection{Platform Ecosystem Diversity of Data Sources}\label{2.2}
The ultimate goal of the detection and governance of online toxic comments is to serve specific online communities and foster a clean and healthy platform communication environment. However, unlike foreign literature that mostly takes Twitter, Reddit, and other platforms as research objects, the Chinese Internet is not a homogeneous entity but a complex ecosystem composed of diverse platforms with significant differences in user profiles, content forms, and community cultures. There are distinct differences in the core functions, user group characteristics, community cultures, and content forms among different platforms, which leads to distinct cross-platform differences in the manifestation, high-frequency topics, and linguistic styles of toxic comments.

Such platform diversity poses severe challenges to toxic comment detection models: a model effective on Weibo may be completely ineffective in Douban Groups; explicit abuse on Tieba may be transformed into veiled female competition rhetoric on Xiaohongshu. Therefore, only by deeply understanding the ecological characteristics of major platforms can we construct a detection system with practical applicability.

\begin{table*}[htbp]
	\centering
	\caption{Summary of Toxic Comment Characteristics Across Mainstream Chinese Online Platforms}
	\label{tab2}
	\renewcommand{\arraystretch}{1.2}
	\begin{tabular}{|m{1.5cm}|m{1.2cm}|m{6.5cm}|m{6.9cm}|}
		\hline
		\textbf{Platform} & \textbf{Severity} & \textbf{User Profile} & \textbf{Key Toxic Characteristics}  \\
		\hline
		Baidu Tieba & {\color{orange}$\bigstar$}{\color{orange}$\bigstar$}{\color{orange}$\bigstar$}{\color{orange}$\bigstar$}{\color{orange}$\bigstar$} & Adolescents and sinking-market users account for over 60\% of the total, with users gathering in interest-based circles. & Extreme virulent invective, indiscriminate abusive remarks, organized brigading attacks on forums, and sustained harassment.  \\
		\hline
		Weibo & {\color{orange}$\bigstar$}{\color{orange}$\bigstar$}{\color{orange}$\bigstar$}{\color{orange}$\bigstar$}{\color{orange}$\bigstar$} & All-age coverage, prominent influence of key opinion leaders, and fan circle users accounting for approximately 18\%. & Rapid spread of collective attacks, doxxing, regional discrimination, rumor dissemination, and fan circle conflicts. \\
		\hline
		Xiaohongshu & {\color{orange}$\bigstar$}{\color{orange}$\bigstar$}{\color{orange}$\bigstar$}{\color{orange}$\bigstar$} & Females account for 72\% of users, with the 18-35 age group as the main demographic, and users exhibit strong demand for consumption decision-making support. & Veiled female competition remarks, appearance and body shaming, incitement of gender/intergenerational confrontation, and subtle emotional abuse. \\
		\hline
		Kuaishou & {\color{orange}$\bigstar$}{\color{orange}$\bigstar$}{\color{orange}$\bigstar$}{\color{orange}$\bigstar$} &  Users from the sinking market account for over 60\%, with a high proportion of users aged 40 and above. & Explicit vulgar jokes, malicious teasing targeting women and the elderly, and remarks involving privacy infringement.   \\
		\hline
		Kuaishou & {\color{orange}$\bigstar$}{\color{orange}$\bigstar$}{\color{orange}$\bigstar$}{\color{orange}$\bigstar$} &  Users from the sinking market account for over 60\%, with a high proportion of users aged 40 and above. & Explicit vulgar jokes, malicious teasing targeting women and the elderly, and remarks involving privacy infringement.   \\
		\hline
		Douban & 
		{\color{orange}$\bigstar$}{\color{orange}$\bigstar$}{\color{orange}$\bigstar$}{\color{orange}$\bigstar$} & Dominantly composed of well-educated literary and artistic youth, with females accounting for 68\% and strong circle cohesion. & Caustic exclusivism within circles, literary-style personal attacks, malicious negative reviews on film and television ratings, and public criticism by posting personal information.  \\
		\hline
		Hupu & 
		{\color{orange}$\bigstar$}{\color{orange}$\bigstar$}{\color{orange}$\bigstar$} & Males account for 89\%, dominated by 20-35-year-old sports/e-sports enthusiasts & Caustic mutual mockery between camps, collective suppression of dissenting opinions, sports fan-style aggressive jokes  \\
		\hline
		Toutiao & 
		{\color{orange}$\bigstar$}{\color{orange}$\bigstar$}{\color{orange}$\bigstar$} & High proportion of sinking-market and middle-aged/elderly users, with strong algorithm dependence & Labeled regional/gender attacks, emotional curses based on news clips, quarrels derived from rumors  \\
		\hline
		Dianping & 
		{\color{orange}$\bigstar$}{\color{orange}$\bigstar$}{\color{orange}$\bigstar$} & Dominantly urban consumer groups, with 25-45-year-olds accounting for over 70\% & Professional negative review extortion, malicious review bombing in peer competition, exaggerated false negative reviews \\
		\hline
		Douyin & 
		{\color{orange}$\bigstar$}{\color{orange}$\bigstar$}{\color{orange}$\bigstar$} & All-age coverage, with average daily usage duration exceeding 120 minutes & Bandwagon-style personal attacks, collective moral judgment in hot events, massive mockery of the parties involved \\
		\hline
		Bilibili & 
		{\color{orange}$\bigstar$}{\color{orange}$\bigstar$} & Dominantly Gen Z (86\% under 35), with strong subcultural identity & Exclusive attacks using memes and circle jargon, favoring insiders and attacking outsiders between different subcultural camps \\
		\hline
		WeChat Channels & 
		{\color{orange}$\bigstar$}{\color{orange}$\bigstar$} & Extension of WeChat ecosystem, high proportion of middle-aged/elderly users, strong acquaintance social attributes & Sporadic intergenerational concept conflicts, stance disputes under hot videos, bickering under toxic chicken soup content \\
		\hline
		Zhihu & 
		{\color{orange}$\bigstar$}{\color{orange}$\bigstar$} & 73\% hold bachelor's degree or above, gathering professionals in specific fields & Logical suppression and superiority-based belittlement under the cloak of rationality, circle-based stance debates and mockery \\
		\hline
		QQ/NetEase Cloud Music & 
		{\color{orange}$\bigstar$} & Dominantly young music lovers, with concentrated pop/rap fans & Occasional disparagement between fan groups, extreme belittlement of singers or work styles \\
		\hline
	\end{tabular}
\end{table*}

To clearly present the core differences among major platforms in dimensions related to toxic comments, Table \ref{tab2} summarizes the severity of toxicity, user profile distribution, and typical toxic characteristics of different platforms, intuitively reflecting the ecological fundamentals of each platform. Based on a comprehensive analysis of platform attributes, public governance reports, and typical public opinion cases, the overall severity of toxic comments across various platforms can be roughly ranked as follows.

\noindent\textbf{1. Baidu Tieba}: A hotspot under anonymity and circle management. As a traditional forum based on interest groups, it is one of the most concentrated and extreme venues for toxic comments in the Chinese Internet due to loose platform supervision and bar owner autonomy. Malicious insults and indiscriminate personal attacks are prevalent in highly emotional topic bars (e.g., sports, games). Its unique "bar brigading" (organized influx to post offensive content) typifies systematic collective cyberbullying, exposing regular users to sustained harassment.

\noindent\textbf{2. Weibo}: Polarization and violence spread in the public opinion sphere. As a core public opinion platform, its toxic comments feature fast transmission, wide impact and strong targeting. Regional discrimination, personal attacks and disinformation thrive amid hot social events, entertainment fan circles and public issues. The ecosystem fosters organized cyberbullying (e.g., fan conflicts, doxxing), with large-scale disputes between users of different stances often spilling offline to cause substantial harm.

\noindent\textbf{3. Xiaohongshu}: Exquisite opposition and anxiety peddling driven by algorithms. Toxic comments are embedded in lifestyle sharing and consumerist discourse. Algorithm-driven recommendations amplify emotional content (e.g., gender relations, in-law conflicts), spawning extreme remarks. Beauty and fashion sections see female competition rhetoric and body shaming rooted in appearance comparisons. Mass-produced confrontational content by MCNs worsens the atmosphere, fostering subtle yet harmful emotional abuse.

\noindent\textbf{4. Kuaishou}: Explicit vulgar attacks in the sinking-market context. Its comment sections feature direct, crude toxicity—prevalent vulgar jokes, malicious teasing of women/elderly, and privacy-infringing insults. Some creators lure traffic by prompting minors to post uncivil content. Despite ongoing platform regulation, the large user base and rapid content production pose persistent challenges to cleaning the comment environment.

\noindent\textbf{5. Douban}: Caustic discipline and exclusive attacks in circle-based communities. Toxicity is tied to its closed interest group culture. Film/TV rating sections have an industrial chain of malicious review bombing; internal group conflicts over opinions, regions or hobbies trigger sharp exclusive verbal attacks. Gender and marriage topics see irrational personal attacks, reflecting intense discipline and conflicts within small circles.

\noindent\textbf{6. Hupu}: Normalization of confrontational discourse in sports communities. A male-dominated sports platform, its comment culture advocates blunt, even aggressive expression. Satire, belittlement and personal attacks on athletes, teams or dissenting users are nearly normalized in sports/e-sports sections. The downvote mechanism reinforces group suppression of dissent, eroding rational discussion space and making provocative remarks hard to eradicate.

\noindent\textbf{7. Toutiao}: Labeled emotional outbursts under hot news. An algorithm-driven news aggregator, its toxic comments concentrate in social news sections. Negative events trigger biased attacks on regional groups; controversial topics (e.g., gender, policies) devolve into labeled malicious curses and personal attacks. Low-quality marketing accounts spread disinformation, escalating confrontations.

\noindent\textbf{8. Dianping}: Fake malicious reviews driven by commercial interests. Toxic comments are overtly utilitarian—professional reviewers extort merchants, and competitors hire online army for fake negative reviews. Some users post exaggerated/fabricated bad reviews (e.g., false food safety claims) over minor grievances, harming merchants' reputation and interests. The mix of true and fake content makes identification and regulation challenging.

\noindent\textbf{9. Douyin:} Bandwagon-style group violence with a huge user base. Despite a diverse content ecosystem, its massive user base and short videos' high emotional contagion make it a hotbed for bandwagon cyberbullying. Hot events (e.g., major competitions, social news) trigger large-scale one-sided personal attacks and moral trials. Intensive platform regulation still faces pressure from instant massive inappropriate comments.

\noindent\textbf{10. Bilibili}: Meme-based attacks and clique favoritism in subcultural circles. Toxic comments are deeply tied to its unique subculture. Exclusive attacks using memes and jargon are rising in games/anime sections; organized disputes (even doxxing) occur between user factions (e.g., game fan groups). Toxic remarks, wrapped in community-specific expressions, are highly concealed to outsiders.

\noindent\textbf{11. WeChat Channels}: Sporadic conflicts extending from acquaintance social networks. As an extension of WeChat's ecosystem, toxicity is relatively scattered. Generational content differences trigger young users' mockery of elders' posts; regional/gender confrontations emerge under hot videos. However, the lack of centralized public forums and constraint of social networks prevent sustained large-scale toxic comment waves.

\noindent\textbf{12. Zhihu:} Stance debates and superiority-based attacks under the guise of rationality. A knowledge-sharing community, its toxic comments are more concealed—logical suppression, sarcasm, or superiority-driven belittlement in debates. Fan conflicts exist in entertainment/game circles, but extreme vulgar remarks are quickly restrained by community rules.

\noindent\textbf{13. QQ/NetEase Cloud Music}: Low-toxicity environments in vertical scenarios. Pure music streaming platforms, their comment sections focus on music itself. Toxicity only occasionally occurs as disparagement between celebrity fan groups. The single-use scenario and strict platform rules make their comment environments the most benign among the surveyed platforms.

Core Implications of Platform Ecology for NLP Detection: Previous sections have clarified the overarching concept of toxicity and its hierarchical terminology framework, while revealing the heterogeneity of the Chinese platform ecosystem, these factors lead to distinct platform-specific characteristics of toxic comments in expression and behavior patterns. Together, they indicate that toxic comment detection should neither conduct generalized identification divorced from the unified toxicity framework nor adopt one-size-fits-all solutions ignoring platform differences; it must align with dual requirements of conceptual hierarchy and platform adaptation. This core orientation serves as the fundamental starting point for dataset construction and detection model advancement in subsequent sections.

\section{Current Status and Methods of Chinese Toxic Comment Dataset Construction}\label{3}
High-quality datasets are the core cornerstone of Chinese toxic comment detection research. This section aims to systematically sort out the construction logic and methods of datasets. First, by critically reviewing existing public datasets, their core limitations, including subjective annotation, coarse granularity, and vague concepts, are clarified. To address these issues, a progressive solution will be proposed subsequently: \ref{3.2} will put forward a novel fine-grained definition and hierarchical theoretical framework, providing clear and unified conceptual basis for data annotation; Based on this framework, \ref{3.3} will explore how to realize efficient, low-cost, large-scale and high-quality annotation with the help of human-machine collaboration, LLMs and other technologies; \ref{3.4} will focus on multi-dimensional quality verification and timeliness evaluation of datasets.
\subsection{Critical Review of Existing Public Datasets} \label{3.1}
Based on the hierarchical framework of toxicity concepts established in Section \ref{2} and insights into the diversity of the Chinese online platform ecosystem, this subsection focuses on the systematic sorting of existing data resources.

\noindent{\textit{(1) Development of Chinese Toxic Comment Datasets}}

The construction history of Chinese toxic comment datasets clearly shows an evolutionary trajectory of research focus from macro to micro and from extensive to refined, which can be divided into three core stages.

The early stage mainly addressed the "from scratch" issue, with datasets mostly built for specific and explicit toxic subcategories. For instance, the NTU Irony Corpus \cite{NTU_Irony} focuses on irony detection, while RPCT \cite{RPCT} and TOCP \cite{TOCP} target profanity. These small-scale datasets adopt a binary classification annotation system, and their core value lies in verifying the feasibility and necessity of constructing dedicated resources for specific Chinese toxic phenomena, providing basic benchmarks and problem orientations for subsequent research.

With the rise of pre-trained language models and the deepened understanding of problem complexity in the research community, the construction of datasets has entered a systematic stage. A landmark achievement of this stage is the proposal of the COLD benchmark \cite{COLD}, which, for the first time, constructs a large-scale, multi-topic, and quality-controllable dataset and evaluation system from the unified perspective of offensive language, a concept closer to the core of toxicity. In the same period, datasets such as SWSR \cite{SWSR} focused on gender discrimination and CDIAL-BIAS \cite{COIAL-Bias} focused on conversational bias have developed in depth, conducting fine-grained and multi-dimensional annotations for specific toxic subcategories. Datasets in this stage begin to emphasize the diversity of data sources and consistency of annotations, striving to become recognized benchmark resources in the field.

Current research frontiers directly address the challenge of Chinese platform ecosystem complexity identified in Section \ref{2}. Dataset construction shows three key trends: first, scenario deepening—e.g., SCCD \cite{SCCD} constructs conversation-level bullying data, and Zhang et al. \cite{Zhang2025} supplement rich contextual information for Bilibili data, accurately responding to the core characteristics of interactivity and context dependence in platform ecosystems; second, adversarial evaluation—perturbation datasets represented by ToxiCloakCN \cite{ToxiCloakCN} and CangjieToxi \cite{CangjieToxi} specifically test model robustness against evasion strategies (e.g., homophone substitution, character splitting), revealing the vulnerability of existing technologies in handling dynamically evolving attacks; third, fine-grained and structured annotation—e.g., STATE ToxiCN \cite{STATE_ToxiCN} requires span-level hate element extraction, advancing detection tasks toward deeper semantic understanding.

\noindent{\textit{(2) Comprehensive Classification and Platform Adaptability Analysis of Existing Datasets}}

To systematically sort out existing data resources in the field of Chinese toxic comment detection, we conduct a comprehensive classification and evaluation of the datasets summarized in Table \ref{tab1}, with the hierarchical framework of toxicity concepts and conclusions on the diversity of the Chinese platform ecosystem (both established in Section \ref{2}) as the core basis. Detailed results are presented in Table \ref{tab3}.

\begin{table*}[htbp]
	\centering
	\caption{Summary of Chinese Toxic Comment Datasets}
	\label{tab3}
	\renewcommand{\arraystretch}{1.2}
	\begin{tabular}{|m{2.5cm}|m{4.3cm}|m{10.2cm}|}
		\hline
		\textbf{Toxicity Category} & \textbf{Representative Datasets}  & \textbf{Adaptation Evaluation} \\
		\hline
		Sarcasm/Irony & NTU Irony \cite{NTU_Irony}, \cite{Gong2020}, CTSD \cite{CTSD}, \cite{Zhang2022}, \cite{Li2020}, CSarc-Context \cite{CSarc-Context}, \cite{Zhu2022}, Observers Corpus \cite{Observers_Corpus}, \cite{Ren2024} & Highly adaptive. It has noticed strong context-dependent scenarios such as news and videos, and has begun to cover platforms like Douyin and Bilibili, accurately capturing their unique contextualized irony patterns. \\
		\hline
		Profanity, Abusive  & RPCT \cite{RPCT}, TOCP \cite{TOCP}  & Low adaptability: Low adaptability. Non-mainland mainstream data sources create a notable gap in wording habits and variants vs. high-frequency profanity on mainland social media. \\
		\hline
		Toxicity, Offensive & COLA \cite{COLA}, COLD \cite{COLD}, AugCOLD \cite{AugCOLD}, TOXICN \cite{TOXICN}, \cite{Zhang2022}, \cite{Li2023}, \cite{Hou2024}, \cite{Cao2024}, ChineseHarm-Bench \cite{ChineseHarm-Bench} &  Moderate adaptability. As a core benchmark, it covers important public discussion platforms, but lacks sufficient depth in covering circle-based and subcultural expressions.   \\
		\hline
		Hate, Bias Expression, Sexism & CDIAL-BIAS \cite{COIAL-Bias}, SWSR \cite{SWSR}, STATE ToxiCN \cite{STATE_ToxiCN}, MCIHD \cite{MCIHD} & Moderate to high adaptability. Coverage of forms on platforms like Xiaohongshu and Hupu is limited, while MCIHD’s attempt at cross-domain coverage marks a significant progress.  \\
		\hline
		Cyberbullying & SCCD \cite{SCCD} &  Highly adaptive and innovative.
		It pioneered session-level annotation, which can accurately capture the repetitive, interactive and progressive harm characteristics of bullying behaviors on platforms like Weibo.
		\\
		\hline
		Various evasion forms of toxicity & ToxiCloakCN \cite{ToxiCloakCN}, CangjieToxi \cite{CangjieToxi}, HED-COLD \cite{HED-COLD}, TE-Dataset \cite{TE-Dataset}
		& Highly forward-looking and adaptive: constructing test sets for evasion tactics is crucial for evaluating model viability in real adversarial environments, yet its general adaptability to cross-platform evasion patterns is insufficient.  \\
		\hline
		Toxicity in Multimodal carriers & MCSh \cite{MCSh}, MCSD \cite{MCSD}
		& Emerging exploration: it begins to fill the gap, yet there remains a significant disparity between its current scale, scenarios and the massive image-text, short-video comment ecosystem of mainstream social media.  \\
		\hline
	\end{tabular}
\end{table*}

\noindent{\textit{(3) Critical Analysis}}

Despite the aforementioned progress, when evaluated against the criterion of building detection systems truly tailored to the Chinese cyberspace, the existing dataset system still has unresolved in-depth contradictions and limitations.

The first is the contradiction between broad platform coverage demand and narrow data representativeness. As shown in Table \ref{tab2}, the Chinese platform ecosystem is highly differentiated in user profiles, community cultures and discourse systems. However, current high-quality datasets are predominantly centered on Weibo and Zhihu, they either lack dedicated collection of platform-specific toxic content (e.g., explicit abuse on Kuaishou's sinking market, subtle emotional abuse on Xiaohongshu, meme-based attacks on Bilibili, extreme anonymous insults on Tieba) or include such content with extremely low proportions in multi-platform datasets. This sampling bias makes models prone to becoming Weibo/Zhihu toxic comment experts, facing severe domain adaptation issues on other platforms, with their generalization ability significantly overestimated.

The second contradiction lies between the academic rigor of annotation systems and the dynamics of online semantics. Most existing datasets ensure static annotation consistency via multi-annotator verification and Kappa coefficient validation. However, online semantic meaning is inherently fluid and consensus-based: the toxic connotations of a single term (e.g., \begin{CJK}{UTF8}{gbsn} 小仙女 \end{CJK} [little fairy], \begin{CJK}{UTF8}{gbsn} 普信男 \end{CJK} [average-looking and confident man]), meme, or emoji change drastically with time, communities, and contextual discussions—an issue that current static, closed annotation systems fail to address. For instance, a homophonic meme labeled neutral last year may have evolved into an offensive code among specific groups this year. The dataset construction cycle lags far behind the evolution speed of online semantics, directly rendering the knowledge learned by models time-limited.

The third contradiction is between the ideal of fine-grained annotation and the reality of subjective, high-cost annotation. While datasets like STATE ToxiCN \cite{STATE_ToxiCN} demonstrate the potential of fine-grained structured annotation, their extension to all types of toxic content and platforms faces significant barriers: judging implicit bias and malicious irony relies on annotators' cultural backgrounds and linguistic intuition, making subjectivity unavoidable; additionally, the high cost of session-level (e.g., SCCD \cite{SCCD}) and span-level annotation severely limits data scale. This directly leads to a data ecosystem gap: on one end are small-scale, high-quality ivory tower datasets; on the other are million-scale augmented datasets (e.g., AugCOLD \cite{AugCOLD}) with high noise and coarse annotations. How to produce large-scale, high-quality annotated data in a low-cost and sustainable manner remains a core challenge in this field.

The fourth contradiction is between the revelation of model vulnerability and the lack of systematic repair solutions. Perturbation datasets such as ToxiCloakCN \cite{ToxiCloakCN} have successfully completed stress tests, exposing models' significant vulnerability to simple evasion strategies. However, these datasets currently serve only as evaluation tools rather than training resources: while they point out the direction for robustness optimization, large-scale training datasets with rich adversarial examples, capable of systematically enhancing models' anti-interference capabilities, have not yet been developed. Further efforts are needed to build an offense-defense integrated data ecosystem.

\subsection{Fine-Grained Definition, Grading, and Correlation of Toxic Comments}\label{3.2}
One of the core cruxes of annotation biases in existing datasets and insufficient model generalization lies in the lack of a unified, clear, and operable definition standard for toxicity. Simple binary classification not only obscures the inherent rich gradient characteristics of toxicity but also fails to adapt to the diverse expression forms in the Chinese cyberspace. To address this, this review breaks through the traditional coarse-grained framework and constructs a multi-dimensional fine-grained definition and grading system integrating type, intensity, target, and intent. This system not only provides a precise benchmark for high-consistency data annotation but also lays a theoretical foundation for models to perform fine-grained understanding and graded disposal tasks, and ultimately realize cross-platform differentiated content governance.

\noindent\textit{(1) Multi-Dimensional Grading Index System}

To precisely characterize the core features of toxic comments, we proposes a grading framework consisting of four relatively independent dimensions. Each dimension focuses on a distinct aspect of toxicity, jointly forming a comprehensive attribute profile (see Table \ref{tab4} for details). Notably, non-toxic comments have clear judgment criteria and thus are not discussed in this study.

\begin{table*}[htbp]
	\centering
	\caption{Multi-Dimensional Grading Annotation System for Toxic Comments}
	\label{tab4}
	\renewcommand{\arraystretch}{1.2}
	\begin{tabular}{|m{1.3cm}|m{4.1cm}|m{7.6cm}|m{3.2cm}|}
		\hline
		\textbf{Dimension}  & \textbf{Grading Description} & \textbf{Annotation Guidelines} & \textbf{Associated Concepts} \\
		\hline
		A.Type & A1.Personal Attack/Profanity\newline
		A2.Discrimination \& Prejudice \newline
		A3.Threat \& Incitement \newline
		A4.Malicious Denigration
		& For comments with multiple types, prioritize the type with the highest proportion and most obvious intent. Example: \begin{CJK}{UTF8}{gbsn}某类人又蠢又坏，别跟他们接触 \end{CJK} [People of a certain kind are stupid and bad; don’t associate with them], core is A2, not A1. & All toxic sub-concepts including prejudiced behavior, toxic irony, and cyberbullying.\\
		\hline
		B.Intensity  & 
		B1.Implicit/Euphemistic\newline
		B2.Direct Denigration\newline
		B3.Extremely Vicious
		& Focus on the explicitness of language and intensity of offensive words. Examples: \begin{CJK}{UTF8}{gbsn}真下头\end{CJK}[Really annoying] (B1), \begin{CJK}{UTF8}{gbsn}你真垃圾\end{CJK} [You’re such trash] (B2), \begin{CJK}{UTF8}{gbsn}祝你早点死\end{CJK} [I hope you die soon] (B3).
		 & Toxic irony (mostly B1), cyberbullying (mostly B2/B3)\\
		\hline
		C.Target & 
		C1.Specific Individual\newline
		C2.Specific Group\newline
		C3.Indiscriminate Attack
		& Identify the scope of the attacked party. Group-targeted toxicity (C2) usually causes broader social harm; clarify group characteristics (e.g., gender, region) as a priority during annotation.
		& Prejudiced behavior (mostly C2), cyberbullying (mostly C1) \\
		\hline
		C.Intent & 
		D1.Emotional Venting\newline
		D2.Belittlement \& Humiliation\newline
		D3.Inciting Confrontation \& Harm
		& Judge based on explicitly expressed intent in the text. Examples: \begin{CJK}{UTF8}{gbsn} 你真烦 \end{CJK}[You’re so annoying] during an argument (D1), sarcastic \begin{CJK}{UTF8}{gbsn} 你真行啊 \end{CJK} [You’re really something] (D2).
		& Systemic discrimination (mostly D3), cyberbullying (mostly D2) \\
		\hline
	\end{tabular}
\end{table*}

Within the framework of the toxic meta-concept established in Section \ref{2}, strict operational definitions of the key sub-concepts and their boundaries in Table \ref{tab4} are required to eliminate ambiguous areas in annotation.

Prejudiced Attitude, Prejudiced Behavior, and Systemic Discrimination:
\begin{itemize}
	\item Prejudiced Attitude: An inherent negative cognitive tendency (e.g., \begin{CJK}{UTF8}{gbsn}我不喜欢某类人\end{CJK} [I don’t like a certain group of people]) that is not externalized into words or actions. It is not classified as a toxic comment, has no corresponding grading indicators, and is directly excluded from annotation.
	\item Prejudiced Behavior: Prejudice externalized as denigration or stigmatization of a group/individual (e.g., \begin{CJK}{UTF8}{gbsn}XX地方的人素质真不行\end{CJK} [People from XX area have poor quality]). It is a toxic comment, core-corresponding to Grading Indicator A2 (Discrimination \& Prejudice).
	\item Systemic Discrimination: Advocacy of injustice and exclusion against specific groups (e.g., \begin{CJK}{UTF8}{gbsn}公司不该招XX地方的人\end{CJK}  [Companies shouldn’t hire people from XX area]). It is a high-risk toxic comment, corresponding to Grading Indicators A2 (Discrimination \& Prejudice) + D3 (Inciting Confrontation \& Harm).
	\item Correlation: Prejudiced behavior is a common component of hate speech; advocating or threatening violence driven by prejudice constitutes the most extreme form of hate speech.
\end{itemize}

Irony and Toxic Irony:
\begin{itemize}
	\item Irony: A rhetorical device with reversed semantics (e.g., between friends: \begin{CJK}{UTF8}{gbsn}你可真聪明，把杯子摔了\end{CJK} [You’re so smart—you dropped the cup]). It is non-malicious, not classified as a toxic comment, has no corresponding grading indicators, and is directly excluded.
	\item Toxic Irony: A sarcastic comment using irony as a tool to belittle and humiliate the target (e.g., \begin{CJK}{UTF8}{gbsn}姐姐好会买平替，质感一眼就看出来呢\end{CJK} [Sis is so good at buying substitutes—you can tell the texture at a glance]). It is a toxic comment, core-corresponding to Grading Indicators A4 (Malicious Denigration) + B1 (Implicit/Euphemistic).
\end{itemize}

General Insult and Cyberbullying:
\begin{itemize}
	\item General Insult: Occasional direct insult (e.g., \begin{CJK}{UTF8}{gbsn}你真傻\end{CJK} [You’re so stupid]) without a sustained target. It is a basic toxic comment, corresponding to Grading Indicators A1 (Personal Attack/Profanity) + B2 (Direct Denigration) + D1 (Emotional Venting).
	\item Cyberbullying: Sustained attacks based on power imbalance (e.g., repeatedly insulting the same person, exposing their privacy). It is a high-risk toxic comment, corresponding to Grading Indicators A1 (Personal Attack/Profanity) + B2/B3 (Direct/Extremely Vicious Intensity) + C1 (Specific Individual) + D2 (Belittlement \& Humiliation).
\end{itemize}

\noindent\textit{(2) Comprehensive Quantification and Correlation Rules for Toxicity Level}

A single dimension is insufficient to assess the overall toxicity level. we proposes a three-stage comprehensive quantification rule, converting multi-dimensional annotations into four overall toxicity grades (High, Medium, Low, Non-Toxic), with emphasis on its deep correlation with the platform ecosystem.

\textbf{Stage 1}: Centered on A.type and B.intensity. The core basis is: attack type determines the nature of toxicity (whether it crosses high-risk boundaries), and attack intensity determines direct degree of harm. Together they form the core hazard potential of toxicity, with higher priority than target and intent. The core hazards of the following two types of combinations have crossed legal red lines or caused severe psychological harm. A total of 54 combinations require no subsequent revisions and are directly classified as High Toxicity.
\begin{itemize}
	\item Including A3 (Threat \& Incitement): The essence of threat is creating personal/rights risks, and the core of incitement is expanding harm scope. Regardless of expression intensity, target, or intent, they all pose high social harm. Example: A3+B1+C1+D1 \begin{CJK}{UTF8}{gbsn}我会让你付出代价\end{CJK} [I will make you pay for it] – implicit but still a threat.
	\item Including B3 (Extremely Vicious) but excluding A3: The core of extreme viciousness (e.g., insult, curse) is the maximization of subjective malice, directly causing severe psychological harm. Example: A1+B3+C3+D1 \begin{CJK}{UTF8}{gbsn}所有人都该去死\end{CJK} [Everyone deserves to die] – no specific target but extremely vicious, with harm far exceeding ordinary denigration.
\end{itemize}
Whereas for comments with attack types A1/A2/A4 and attack intensity B1/B2, their core harm does not meet the high-toxicity threshold and requires further fine-tuning based on attack target + discernible intent. The initial baseline level is set in accordance with the principle of intensity as follows:
\begin{itemize}
	\item B1 (Implicit/Euphemistic): Indirect expression, ambiguous malice, and weak perceived harm $\to$ Initial baseline: Low Toxicity;
	\item B2 (Direct Denigration): Explicit expression, straightforward malice, and strong perceived harm $\to$ Initial baseline: Medium Toxicity.
\end{itemize}

\textbf{Stage 2}: For the aforementioned 54 to-be-revised combinations, the core basis is: attack target determines the scope of harm (individual $\to$ group $\to$ indiscriminate; the broader the scope, the greater the harm), and discernible intent determines the degree of subjective malice (emotional venting $\to$ belittlement \& humiliation $\to$ inciting confrontation; the stronger the malice, the greater the harm). The two jointly fine-tune the initial baseline level to achieve accurate alignment with the actual harm. Specific fine-tuning settings are shown in Table \ref{tab5}.

\begin{table*}[htbp]
	\centering
	\caption{Fine-Tuning Rules Based on Attack Target and Intent.}
	\label{tab5}
	\renewcommand{\arraystretch}{1.2}
	\begin{tabular}{|m{3.3cm}|m{2.3cm}|m{10.8cm}|}
		\hline
		\textbf{A+B Basic Combination} & \textbf{Fine-Tuning Rules}  & \textbf{Grading Basis} \\
		\hline
		A1+B1 \newline(Initial Low Toxicity) & 1. C2+D2/D3$\nearrow$ \newline 2. Others$\leftrightsquigarrow$ & C2+D2/D3 $\to$ expanded harm scope and increased subjective malice. Example: \begin{CJK}{UTF8}{gbsn}XX玩家真厉害, 只会坑人\end{CJK} [Players from XX team are so skilled—they only ruin the game] $\to$ implicit humiliation against a specific group, with greater harm than individual emotional venting.
		 \\
		\hline
		A1+B2 \newline(Initial Medium Toxicity)  & 1. C2+D3$\nearrow$ \newline  2. C1+D1$\searrow$ \newline 3. Others$\leftrightsquigarrow$ &  C2+D3$\to$Both the scope of harm and malicious intent are maximized (e.g., A1+B2+C2+D3: \begin{CJK}{UTF8}{gbsn}大家别跟 XX 玩家组队, 都是坑\end{CJK} [Everyone should avoid teaming up with Player XX—they are all incompetent] $\to$ incites group exclusion and is classified as high toxicity).\\
		\hline
		A2+B1 \newline(Initial Low Toxicity)  & 1. C2+D2/D3$\nearrow$ \newline  2. C3+D3$\nearrow$ \newline 3. Others$\leftrightsquigarrow$ & The core harm of A2 lies in undermining social equity. C2 or C3 + D3 $\to$ expands the impact of discrimination (e.g., A2+B1+C3+D3: \begin{CJK}{UTF8}{gbsn}某些地方的人都不靠谱, 大家别信\end{CJK} [People from certain regions are all unreliable—no one should trust them] $\to$ generalizes discrimination + incites others, and is classified as Medium Toxicity). \\
		\hline
		A2+B2 \newline(Initial Medium Toxicity)  & 1. C2+D3$\nearrow$ \newline  2. C1+D1$\searrow$ \newline 3. Others$\leftrightsquigarrow$ & e.g., A2+B2+C2+D3: \begin{CJK}{UTF8}{gbsn}XX性别不适合做这个, 大家抵制\end{CJK} [Individuals of gender XX are unfit for this—everyone should resist it] $\to$ discrimination + incitement to antagonism $\to$ high social harm, and is classified as High Toxicity. \\
		\hline
		A4+B1 \newline(Initial Low Toxicity)  & 1. C2+D2$\nearrow$ \newline  2. Others$\leftrightsquigarrow$ & C2+D2 $\to$ Public humiliation within the group, requiring escalation. In other cases, the malicious intent is weak (e.g., C1+D1: \begin{CJK}{UTF8}{gbsn}你这衣服好像不太值\end{CJK} [Your clothes seem not quite worth the price] $\to$ merely individual emotional release, and is classified as slightly toxic). \\
		\hline
		A4+B2 \newline(Initial Low Toxicity)  & 1. C2+D2/D3$\nearrow$ \newline  2. C1+D1$\searrow$ \newline 3. Others$\leftrightsquigarrow$ & C2+D2/D3 $\to$ e.g., \begin{CJK}{UTF8}{gbsn}买平替的都是爱占便宜, 大家别学\end{CJK} [Those who buy substitutes are all greedy for petty gains—no one should follow suit] $\to$ incites group exclusion and is upgraded; C1+D1 $\to$ individual emotional release without malicious intent, and is downgraded\\
		\hline
	\end{tabular}
\end{table*}

\textbf{Stage 3}: Platform Context Calibration. Toxicity perception of the same comment varies with platform community culture and communication norms; thus, grading results should align with users' actual experience to avoid over- or under-regulation caused by a one-size-fits-all approach. Calibration does not alter the core judgment of highly toxic content (its harm has reached the threshold, requiring unified cross-platform regulation) but only fine-tunes medium/low-toxic combinations.

Taking Bilibili as an example, for the combination A2+B2+C1+D2, according to the unified rules of Stage 2, the initial benchmark of A2+B2 is medium toxicity. Since this combination meets neither the upgrading criteria of C2+D3 nor the downgrading criteria of C1+D1, it is judged as medium toxicity before calibration. The corresponding comment is: \begin{CJK}{UTF8}{gbsn}某UP主这波操作纯纯提纯, 真是没脑子\end{CJK} [This move by a certain UP owner is totally purification—really brainless]. Here, \begin{CJK}{UTF8}{gbsn}提纯\end{CJK}[purification] is a fan circle jargon on Bilibili, actually used by users to complain about the UP owner’s improper operation without discriminatory or humiliating intent. Bilibili has intensive circle cultures, where jargon and meme-based expressions are core communication methods. In the fan circle context, such remarks are mostly mild complaints within the circle; users’ primary perception is dissatisfaction with the operation rather than personal attack, and the actual malicious perception is much lower than the judgment based on unified rules. Thus, it is downgraded from medium toxicity to low toxicity after calibration, avoiding over-restricting normal communication in Bilibili’s circles while retaining the identification of implicit improper expressions.

Taking Xiaohongshu as an example, for the combination A2+B1+C2+D1, according to the unified rules of Stage 2, the initial benchmark of A2+B1 is low toxicity. Since this combination does not meet the upgrading criteria, it is judged as low toxicity before calibration. The corresponding comment is: \begin{CJK}{UTF8}{gbsn}XX职业的人来做美妆分享, 果然还是差点意思呢\end{CJK} [People in XX profession doing beauty sharing are really not quite up to par]. Though without explicit derogatory terms, it implies discrimination that people in this profession lack the qualifications for beauty sharing and targets a specific group. Xiaohongshu focuses on scenario-based sharing with a high proportion of female users, where users are highly sensitive to implicit discrimination and group exclusion—such euphemistic expressions make users in the corresponding profession feel negated and excluded, with actual malicious perception far exceeding the low toxicity judgment. Thus, it is upgraded from low toxicity to medium toxicity after calibration, aligning with the actual harm perception of platform users while upholding the governance bottom line against discrimination.

Without platform context calibration, merely mechanically applying the unified results of Stage 2 will lead to misjudgments: the jargon-based complaints on Bilibili will be incorrectly classified as medium toxic, excessively restricting circle communication; the implicit discrimination on Xiaohongshu will be misjudged as low toxicity, condoning exclusion of specific groups. Such annotation bias divorced from platform context will cause significant platform heterogeneity errors in the dataset, making the model learn unified rules inconsistent with real scenarios rather than cross-platform universal and platform-adaptive toxicity identification capabilities. Ultimately, this results in a surge in the misjudgment rate of the model in practical applications.

It should be noted that the development of the grading framework inherently involves subjectivity. Although the grading framework designed in this review cannot be guaranteed to be fully scientific, it can facilitate the further formulation of comprehensive quantitative rules covering all toxic comments.

\subsection{Summary of Efficiency-Enhancing Annotation Methods}\label{3.3}
Against the backdrop of the fine-grained annotation framework proposed earlier (4 attack types $\times$ 3 intensity levels $\times$ 3 targets $\times$ 3 intents) and the increased judgment complexity caused by cross-platform new online memes and implicit expressions, the traditional manual-only annotation mode has hit a bottleneck in terms of cost and scalability. This subsection summarizes and prospects the new-generation efficient annotation methodology, whose core paradigm has shifted from sequential manual annotation to human-machine collaboration: human experts define rules and standards, and control quality and ethics, while machine learning models undertake repetitive tasks such as preliminary screening, pre-annotation and generative data augmentation. This achieves an order-of-magnitude improvement in annotation efficiency while ensuring quality.

\noindent\textit{(1) Closed-Loop Workflow of Human-Machine Collaborative Annotation}

\begin{figure}[t]
	\centerline{\includegraphics[width=3.5in,keepaspectratio]{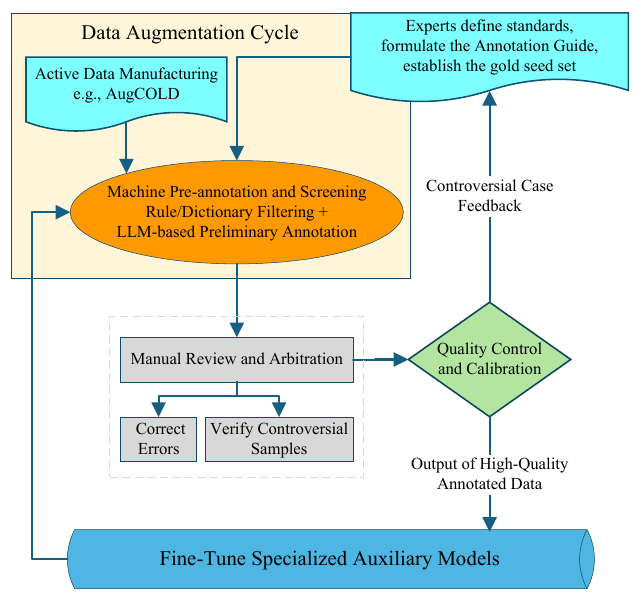}}
	\caption{Closed-Loop Workflow of Human-Machine Collaborative Annotation.}
	\label{fig2}
\end{figure}

A systematic modern annotation pipeline is a dynamically enhanced closed-loop system, with its core workflow illustrated in Fig. \ref{fig2}. The process starts with human experts formulating a detailed Annotation Guide and establishing a small-scale, high-quality gold seed set based on the predefined fine-grained framework, injecting authoritative judgment criteria into the entire system.

Subsequently, machines undertake heavy preprocessing tasks: on the one hand, rapid preliminary screening is performed via keywords, regular expressions, or specialized dictionaries such as SexHateLex \cite{SWSR}, combined with platform-specific feature libraries (e.g., Bilibili jargon dictionary, Xiaohongshu implicit expression annotation library) to improve the target data density of the to-be-annotated set; on the other hand, LLM is used to conduct multi-dimensional pre-annotation with confidence scores for the screened texts, achieving preliminary understanding of massive corpora.

Then, the role of human annotators has shifted from original sequential judges to efficient reviewers and arbitrators. They no longer need to process a large number of simple or irrelevant samples; instead, they concentrate their expertise on reviewing the machine’s pre-annotation results, focusing on tackling difficult cases characterized by low confidence, logical contradictions between dimensions, or cross-platform complex cultural argot (e.g., confrontational context on Tieba, euphemistic expressions on Xiaohongshu). Of particular importance is that human experts systematically verify samples where the machine may make high-confidence erroneous judgments, especially cases of implicit toxicity, and feed these boundary disputes back to the Annotation Guide, enabling continuous iteration and calibration of annotation standards.

Eventually, the high-quality data confirmed by humans is used to fine-tune a specialized annotation-aided model, enhancing its accuracy in the next round of annotation and thus forming a continuously self-strengthening data flywheel. Additionally, for hard samples scarce in natural data (e.g., specific types of implicit discrimination), the idea of AugCOLD \cite{AugCOLD} can be adopted: under the guidance of the fine-grained framework, generative models are proactively used to directionally synthesize adversarial samples, filling gaps in data distribution at low cost and entering a positive data augmentation cycle.

\noindent\textit{(2) From Data Augmentation to AI Annotators}

In the practice of cutting-edge methods, generative AI and LLMs are playing increasingly critical roles \cite{ChineseHarm-Bench,ToxiCloakCN}. Active data manufacturing methods represented by AugCOLD are valuable in transforming machine learning from passive data consumers to active participants. For instance, to train models to accurately identify implicit derogation against specific occupations, researchers can instruct generative models to batch-generate texts meeting the criteria of "type = discrimination and prejudice, intensity = implicit, target = specific occupational groups, intent = derogation", and simultaneously generate their non-toxic control versions. This enables the rapid construction of a balanced and targeted training set, effectively addressing the shortage of hard samples encountered in previous reliance on web crawlers.

Meanwhile, LLMs have demonstrated enormous potential as annotation capability amplifiers. Through sophisticated prompt engineering and few-shot learning, LLMs can make complex dimensional judgments on texts at a near-human level. In practice, a strategy of independent annotation by multiple LLMs followed by integrated voting can be adopted, which estimates the annotation controversy of each sample and intelligently directs limited manual review resources to the most uncertain and discriminatively needed cases.

However, their limitations must be prudently recognized: LLMs may have delayed understanding of dynamically evolving online memes and in-depth subcultural contexts, especially in cross-platform scenarios where their ability to adapt to the context of different platforms varies; their judgments may inadvertently inherit social biases in pre-training data; their performance remains unstable in dimensions highly dependent on context and subjective understanding, such as intent inference. Therefore, at this stage, LLMs are more appropriately positioned as efficient assistants under strict human supervision and quality control, rather than fully autonomous annotation subjects.

\noindent\textit{(3) Engineering Support: Doccano}

Transforming the aforementioned human-machine collaboration theoretical framework into repeatable and manageable daily operations relies on engineering toolchain support. Open-source annotation platforms, typified by Doccano\footnote{https://github.com/doccano/doccano}, play a pivotal role: project managers can conveniently upload and share the Annotation Guide to ensure unified standards among annotation teams; they support direct import of machine pre-annotation results to enhance manual review efficiency; built-in functions such as multi-person collaboration, task assignment, and highlighting of inconsistent annotations greatly facilitate team collaboration and arbitration discussions; the provided basic data statistics functions help monitor annotation quality and progress throughout the process. For cross-platform annotation needs, Doccano’s custom label function enables flexible configuration of multi-dimensional labels and platform identifiers, aligning the annotation process precisely with the fine-grained framework mentioned earlier and further reducing operational complexity.

\subsection{Data Feature Analysis to Verify Quality/Timeliness}\label{3.4}
After the construction of the dataset, its quality and timeliness need to be verified from multiple dimensions to ensure it can truly and effectively support model development.

\noindent\textit{(1) Internal Consistency}

The internal consistency of data is central to reliability assessment, yet the assessment should go beyond the single Kappa coefficient \cite{Kappa}. This coefficient objectively reflects the true consistency by excluding chance agreement caused by random guesses, and its formula is as follows:
\begin{equation}
	\kappa=\frac{P_o-P_e}{1-P_e},
\end{equation}
where $P_o$ denotes the observed agreement rate, and $P_e$ the expected chance agreement rate. Generally, $\kappa \geq 0.8$ indicates high agreement, while $\kappa < 0.4$ suggests poor consistency.

First, inter-annotator reliability should be calculated separately for each dimension of the fine-grained framework. For example, a dataset may have $\kappa$=0.9 (high agreement) for attack types but $\kappa$=0.5 (low reliability) for intent, which precisely reveals ambiguities in annotation guidelines or concept comprehension for the latter, providing clear direction for standard iteration. Second, intra-annotator reliability testing, where the same annotator re-annotates some samples after an interval, with Kappa coefficient calculated for re-annotation results, can assess the personal-level stability of annotation standards and the clarity of the Annotation Guidelines. Most critically, the association rules established in \ref{3.2} must be used to conduct logical consistency checks.

\noindent\textit{(2) Temporal dynamics analysis}

Internet language evolves rapidly, so the timeliness of datasets is critical. On one hand, lexical novelty analysis should be conducted: count high-frequency toxic keywords in the dataset, compare them with peer datasets or general corpora from 1-2 years prior, and calculate the proportion of new words/memes, a high proportion indicates the dataset captures recent expressions. On the other hand, temporal decay testing of model performance is needed: train a baseline model on the current dataset, then evaluate its performance on time-sequenced test subsets; a significant performance drop on newer time slices strongly signals data distribution drift, requiring dataset updates.

\noindent\textit{(3) Platform Ecology Representativeness Analysis}

Based on the ecological analysis in \ref{2}, the platform coverage bias of the dataset is evaluated. First, the data source distribution is counted, and the proportion of data samples from each platform is visualized and compares with the monthly active users of the platform or the overall traffic share of the Chinese Internet, so as to intuitively determine whether there is over-representation or under-representation. Second, the cross-platform domain adaptation difficulty test poses a more severe challenge: a model trained on data from a single mainstream platform (e.g., Weibo) is directly subjected to zero-shot evaluation on the test set from another platform (e.g., Xiaohongshu). A significant performance gap confirms the domain gap caused by a single data source, highlighting the importance of constructing a cross-platform balanced dataset.

\noindent\textit{(4) Adversarial Robustness Testing}

The ultimate value of a dataset lies in its ability to train models that truly understand semantics rather than merely memorize surface patterns. Using specialized perturbation test sets such as ToxiCloakCN and CangjieToxi, the robustness of models trained on the dataset can be evaluated. The magnitude of performance degradation of models on such test sets is a key indicator of whether they memorize surface patterns instead of truly understanding toxic semantics, and also indirectly reflects whether the original dataset contains sufficient adversarial variants.

An ideal Chinese toxic comment dataset must stand the test across four dimensions: internal consistency, temporal freshness, ecological representativeness, and adversarial robustness. This requires shifting dataset construction from a one-off project to a long-term endeavor involving continuous monitoring, dynamic expansion, and iterative optimization, thus keeping pace with the evolving Chinese online ecosystem and sustaining the vitality of detection models.

\section{Technical Evolution of Chinese Toxic Comment Detection Models}\label{4}
The development of Chinese toxic comment detection technology is a progressive process that advances in tandem with the evolution of NLP paradigms. Its core driving force lies in addressing the increasingly covert and complex forms of toxic expression. Toxic expression has evolved from direct verbal abuse in the early stage to covert expression via homophones, metaphors, and code names \cite{TE-Dataset}. To clearly illustrate this evolutionary trajectory, Table \ref{tab6} summarizes the core ideas, typical methods, technical merits, and inherent limitations of each major technical stage.

\begin{table*}[htbp]
	\centering
	\caption{The Developmental Landscape of Chinese Toxic Comment Detection Models.}
	\label{tab6}
	\renewcommand{\arraystretch}{1.2}
	\begin{tabular}{|m{1.0cm}|m{3.4cm}|m{3.0cm}|m{3.8cm}|m{4.6cm}|}
		\hline
		\textbf{Stage} & \textbf{Core ideas}  & \textbf{Typical Methods} & \textbf{Advantages} & \textbf{Limitations} \\
		\hline
		Rules and Lexicon & String matching based on explicit rules and lexicons & Keyword blacklist, lexicon lookup, pronunciation features & Simple, efficient, highly interpretable, and low in computational cost & 	Incapable of handling expression variants and semantic information, with high rates of missed and false judgments as well as high lexicon maintenance costs
		\\
		\hline
		Statistical ML & Automatically learning classification features and patterns from annotated data & TF-IDF, n-gram combined with SVM/LR; Word2Vec, fastText  & Possesses certain generalization ability and can capture shallow semantic and word order features & Relies on manual feature engineering and fails to comprehend deep semantics and complex contexts
		\\
		\hline
		DL & Using neural networks to automatically learn deep distributed representations of text & CNN, LSTM/BiLSTM (often combined with Word2Vec initialization) & Enables automatic feature extraction and captures long-distance contextual dependencies & Requires large-scale annotated data, with limited capability in recognizing implicit expressions and domain adaptation
		\\
		\hline
		Pre-trained Model & Pre-training on large-scale corpora, followed by fine-tuning to adapt to downstream tasks & Fine-tuning of BERT, RoBERTa and their variants (e.g., bert-base-chinese) & Boasts strong contextual semantic understanding and serves as the current mainstream strong baseline & Demands substantial computational resources and remains sensitive to the quality of annotated data and application domains
		\\
		\hline
		LLM & Leveraging the strong general knowledge and reasoning capabilities to achieve judgment & Prompt engineering based on models such as GPT & Ready-to-use without training, and excels at understanding complex logic and implicit semantics & Suffers from high inference latency, poor result stability, hallucination issues, and exorbitant costs
		\\
		\hline
	\end{tabular}
\end{table*}

\subsection{Overview of Existing Chinese Toxic Comment Detection Models}\label{4.1}

\noindent\textit{(1) Rule and Lexicon-Based Method}

As the earliest and most intuitive detection technique in Chinese toxic comment detection, the rule and lexicon-based method is centered on the core principle of performing string matching on text against pre-defined sensitive lexicons (blacklists) or rule patterns \cite{Hurtlex,Lee,Razavi}. If a match is detected, the text is classified as toxic content. Common implementation approaches include:
\begin{itemize}
	\item Keyword Matching: Directly scans comments for words in a predefined sensitive lexicon. For example, if "\begin{CJK}{UTF8}{gbsn}傻瓜\end{CJK} [fool]" is listed, comments containing it are blocked. This method is simple but easily bypassed by variants like "\begin{CJK}{UTF8}{gbsn}傻X \end{CJK}" or "\begin{CJK}{UTF8}{gbsn}沙雕 \end{CJK}" [both equivalent to fool].
	\item Lexicon Look-up: An extension of keyword matching that employs a more structured lexicon, potentially incorporating part-of-speech or sentiment information, yet remains fundamentally based on string matching.
	\item Phonetic Feature Matching: Designed to counter homophonic substitutions (e.g., "\begin{CJK}{UTF8}{gbsn}煞笔\end{CJK}" [shabi] for "\begin{CJK}{UTF8}{gbsn}傻逼\end{CJK}" [shabi]) by converting text to phonetic notation (e.g., Pinyin) for comparison.
\end{itemize}

Such methods are widely adopted in early studies and entry-level industrial solutions. For instance, the work of \cite{RPCT} proposed 29 sets of detection and rewriting rules integrating keywords, lexicons and phonetic features. Many studies also use rule-based commercial APIs (e.g., Baidu Text Moderation) as baseline comparison schemes \cite{COLD,TOXICN,SCCD,COIAL-Bias}. The open-source sensitive lexicon \cite{COLD} on GitHub, containing 14,000 entries, also serves as a fundamental resource for such methods. However, the limitations of rules are prominent: they fail to identify semantic variations (e.g., synonym substitution, homophonic modification), combinatorial associations (compliant words combining to form inappropriate meanings) and context-dependent expressions. Thus, such schemes are usually only employed as the first line of defense for rapid filtering or as preprocessing modules.

\noindent\textit{(2) Statistical ML Methods}

After rule- and lexicon-based methods proved inadequate for complex linguistic variations, research shifted to an era dominated by statistical ML methods. The core of this phase lies in automatically learning classification patterns from labeled data, evolving primarily along two directions. On one hand, researchers manually convert text into numerical features via feature engineering approaches. For example, TF-IDF is adopted to quantify word importance, with its core idea being that the more frequently a word appears in the current review, while the more prevalent it is across the entire review corpus, the weaker its discriminative power \cite{TF-IDF}. Its calculation formula is as follows
\begin{equation}
	\text{TF-IDF}(t,d)=\text{TF}(t,d)\times \log\frac{N}{\text{DF}(t)+1},
\end{equation}
where $t$ denotes a term, $d$ denotes the current document, and $\text{TF}(t,d)$ is the term frequency of term $t$ in document $d$. $N$ is the total number of documents, and $\text{DF}(t)$ is the number of documents containing term $t$. The extracted features (including TF-IDF and n-grams that capture local word order) are then fed into classifiers such as support vector machines (SVMs) for training.

On the other hand, static word embedding methods are widely adopted to obtain richer semantic representations than handcrafted features. Represented by Word2Vec \cite{Word2Vec} and fastText \cite{fastText}, such methods are first pre-trained on ultra-large-scale unlabeled corpora to learn a fixed, low-dimensional dense vector for each word. Via the Skip-gram or CBOW architectures, Word2Vec enables the model to learn to predict word contexts, thereby bringing semantically similar words close to each other in the vector space. Building on this, fastText introduces further innovations: it regards each word as a set of its subwords (character n-grams), represents the entire word as the sum of subword vectors, and thus achieves better generalization ability for out-of-vocabulary words and morphological variants (e.g., the variant "\begin{CJK}{UTF8}{gbsn}傻逼\end{CJK}" [shabi] derived from the offensive term "\begin{CJK}{UTF8}{gbsn}沙壁\end{CJK}" [sabi]).

In practical applications, all word vectors in a sentence are usually aggregated (e.g., by averaging), and the resulting sentence representation is then fed into the classifier. These statistical learning methods endow the model with preliminary generalization ability, enabling it to identify certain variants not explicitly listed in the lexicon. However, they share a fundamental limitation: both TF-IDF weights and static word embeddings yield fixed, context-independent representations for each word. This means that the model cannot dynamically distinguish the different meanings of words based on specific contexts, which severely limits its ability to understand complex semantics and implicit expressions.

\noindent\textit{(3) DL Methods}

To break through the bottleneck of static representations and shallow features, DL methods emerged as the times require, whose core lies in leveraging neural networks to automatically learn deep, distributed representations of text. A typical workflow starts with initializing the embedding layer using pre-trained Word2Vec word embeddings to convert text into dense vector sequences; it then uses CNN to extract local phrase features such as "\begin{CJK}{UTF8}{gbsn}真恶心\end{CJK} [really disgusting]" \cite{TOCP,Cao2024,SCCD}, or adopts BiLSTM to capture long-distance contextual dependencies \cite{SWSR,TOXICN,MCIHD}, thereby understanding the modifying relationship between "\begin{CJK}{UTF8}{gbsn}你\end{CJK} [you]" and "\begin{CJK}{UTF8}{gbsn}愚蠢\end{CJK} [stupid]".

The introduction of the attention mechanism further enables models to focus on key toxic segments \cite{COLA,TE-Dataset}. This paradigm reduces reliance on handcrafted feature engineering and significantly enhances the ability to model contextual semantics \cite{Hou2024}. However, its performance is heavily dependent on large-scale, high-quality labeled data, and its generalization ability remains limited for toxic expressions outside the training data distribution, especially highly implicit ones.

\noindent\textit{(4) Pre-trained Models}

Given the limitations of DL, pre-trained language models have initiated a new mainstream paradigm. Represented by models such as BERT \cite{BERT} (bert-base-chinese) and RoBERTa \cite{RoBERTa} (roberta-base-chinese), they acquire robust, contextually dynamic word embedding capabilities (i.e., the same word has different vectors in different contexts) via self-supervised pre-training tasks including masked language modeling on massive unlabeled text corpora. For toxic content detection, the fine-tuning strategy is mainly adopted: a classification layer is added on top of the pre-trained model, followed by end-to-end training on specific toxic-labeled datasets to enable rapid adaptation to downstream tasks.

To address the identification challenges posed by implicit expressions (e.g., homophones and euphemisms) in Chinese toxic comments, cutting-edge studies have upgraded the fine-tuning framework via multi-augmentation techniques, yielding a series of breakthrough results. The COLD \cite{COLD} benchmark dataset constructed by Deng et al. in 2022 systematically collected samples of Chinese implicit toxic expressions for the first time, laying a data foundation for subsequent research; their proposed \cite{AugCOLD} in 2023 further integrates data augmentation and prompt tuning, embeds task descriptions into input texts, and effectively stimulates the semantic comprehension potential of pre-trained models. In 2024, Hou et al. \cite{Hou2024} proposed a Chinese offensive language detection algorithm enhanced by a pointer network based on the pre-trained RoBERTa model. The algorithm follows a workflow consisting of text preprocessing, task prompt injection, deep semantic extraction via RoBERTa, focused vector acquisition using a pointer network, and vector fusion for classification. Its detection performance on the COLD dataset significantly outperforms that of mainstream models such as FastText and BERT. Proposed by Zhang et al. in 2024 \cite{MCIHD}, the domain prompt learning method combines domain feature fusion with prompt learning. It outperforms mainstream models such as TF-IDF and BERT-CLS in both few-shot and full-shot scenarios, achieving state-of-the-art performance in Chinese implicit hate speech detection. Collectively, these methods have pushed the performance of pre-trained models in Chinese toxic speech detection to new heights, serving as the core technical support for constructing high-performance detection systems in academia.

\noindent\textit{(5) LLMs}

Meanwhile, the rise of LLMs has spawned a revolutionary assessment-as-a-service paradigm. Centered on trillion-parameter models such as GPT-4o and DeepSeek-V3, these approaches require no specialized training for toxic content detection tasks; instead, researchers can directly activate the models’ internal knowledge to perform zero-shot or few-shot inference via well-designed prompt engineering (e.g., "Act as a content moderation expert to determine whether the text contains offensive content and output only 'Yes' or 'No'").

Relevant evaluation benchmarks (e.g., CangjieToxi \cite{CangjieToxi}, ChineseHarm-Bench \cite{ChineseHarm-Bench}) have systematically verified the task potential of LLMs. Combined with cross-validation from studies such as ToxiCloakCN \cite{ToxiCloakCN}, STATE ToxiCN \cite{STATE_ToxiCN}, SCCD \cite{SCCD} and TED-SCL \cite{TE-Dataset}, these models are found to exhibit outstanding performance in parsing complex contexts, capturing long-text correlations and interpreting the semantics of implicit expressions. However, this paradigm still faces inherent challenges: high inference costs, potential hallucination biases in generated results, and high sensitivity of performance to the wording design of prompts.

\subsection{Interpretability Analysis in Chinese Toxic Comment Detection}\label{4.2}
Interpreting model decision-making mechanisms is not only key to evaluating their reliability and fairness, but also a core component of enabling detection result traceability, guiding feature and rule optimization, and thus constructing a closed-loop improvement system. It is therefore necessary to thoroughly explore the black-box nature and interpretability issues of models across different technical pathways. In what follows, we systematically analyze the progress and limitations of interpretability from rule-based methods to LLMs, by integrating specific research cases and interpretive techniques.

\noindent\textit{(1) White-box Transparency of Rule- and Lexicon-based Methods}

Rule- and lexicon-based methods are inherently fully interpretable white-box systems. Their decision-making logic is deterministic: any judgment result can be directly attributed to one or more triggered keywords, regular expression patterns, or phonetic matching rules. For example, when the system blocks the comment "\begin{CJK}{UTF8}{gbsn}你真是个傻X\end{CJK} [You are such an idiot]", it can automatically generate an explanation: "Triggered the term '\begin{CJK}{UTF8}{gbsn}傻X\end{CJK} [idiot]' in the sensitive word blacklist". Such transparency provides the only direct pathway for preliminary traceability and rapid optimization. By analyzing false positive logs (e.g., misclassifying the legitimate comment "\begin{CJK}{UTF8}{gbsn}快给我解毒\end{CJK} [Hurry up and detox me]" as toxic), operators can accurately locate and revise overly broad rules; by examining false negative cases (e.g., failing to identify the variant "\begin{CJK}{UTF8}{gbsn}沙雕\end{CJK} [silly sand]" for the offensive term), they can immediately expand the lexicon.

However, their interpretability is limited to superficial string matching and fails to capture semantic meanings. They cannot explain why "\begin{CJK}{UTF8}{gbsn}蠢得可爱\end{CJK} [foolishly cute]" and "\begin{CJK}{UTF8}{gbsn}蠢得可恨\end{CJK} [foolishly hateful]" differ drastically in sentiment despite differing by only one word, nor can they handle metaphorical sarcasm such as "\begin{CJK}{UTF8}{gbsn}你这操作真下饭\end{CJK} [Your gameplay is such a joke]", which relies on gaming community common sense. This fundamental limitation of interpretability reveals the upper bound of pure rule-based systems in terms of complex semantic comprehension.

\noindent\textit{(2) Visual Attribution for Statistical ML Models}

As model complexity increases, interpretability needs to be achieved with the aid of specialized technical tools. For statistical learning models based on traditional feature engineering (e.g., TF-IDF, bag-of-words models), their feature representations are confined to the level of term frequency statistics and lack the ability to capture semantic correlations; thus, local approximation explanation frameworks serve as effective probes for deciphering their decision-making logic, among which local interpretable model-agnostic explanations (LIME) is the most widely used.

\begin{figure*}[t]
	\centerline{\includegraphics[width=6.3in,keepaspectratio]{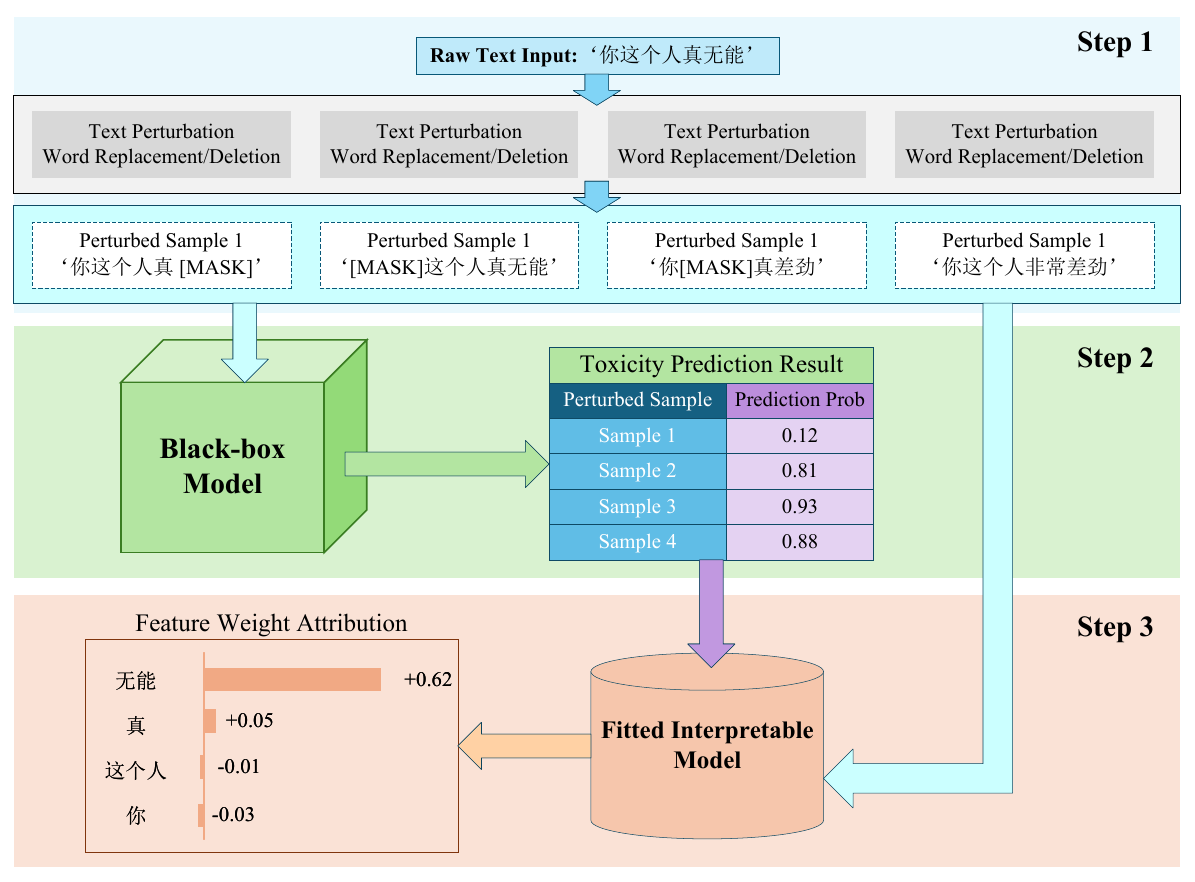}}
	\caption{Schematic diagram of the LIME technical workflow.}
	\label{fig3}
\end{figure*}

The technical logic of LIME can be decomposed into three key steps, as illustrated in Fig. \ref{fig3}. First, for a target input sample (e.g., a text comment to be judged), a set of virtual samples locally similar to the original sample is generated via controlled perturbation methods such as random word replacement and deletion. Second, this sample set is fed into the statistical model to be interpreted (e.g., SVM, logistic regression, etc.) to obtain the corresponding prediction results. Third, using the similarity between virtual samples and the original sample as weights, an interpretable simple model is fitted within the local feature space of the original sample; the contribution of each word to the prediction result is quantified via model coefficients, and the top-ranked key features by weight are finally output. As shown in the results of Fig. \ref{fig3}, for a comment classified as offensive by the model, LIME reveals that the term "\begin{CJK}{UTF8}{gbsn}无能\end{CJK} [incompetent]" has the highest positive weight, which verifies the partial consistency between the model's decision-making basis and human intuition. More importantly, LIME can accurately capture the potential biases of statistical models. For instance, some models may assign relatively high positive weights to neutral degree adverbs such as "\begin{CJK}{UTF8}{gbsn}真\end{CJK} really" and "\begin{CJK}{UTF8}{gbsn}非常\end{CJK} [extremely]", because such adverbs often co-occur with offensive terms in the training data, leading the model to mistakenly take them as judgment criteria.

\noindent\textit{(3) An Analysis of the Attention Mechanism in DL Models}

Compared with statistical models, recurrent neural networks (represented by RNN and LSTM) and DL models integrated with the attention mechanism have achieved semantic-level feature extraction by virtue of their stronger sequence modeling capability, and the interpretability analysis of these models has thus entered the era of visualization. By calculating the contribution weight of each position in the sequence to the final decision, the attention mechanism opens an intuitive window into the black-box decision-making process of the model \cite{RenDCAN}. Such weight distribution can be converted into word-level heatmaps, enabling researchers to directly observe which parts of the text the model focuses on when identifying toxic content.

\begin{figure}[t]
	\centerline{\includegraphics[width=3.6in,keepaspectratio]{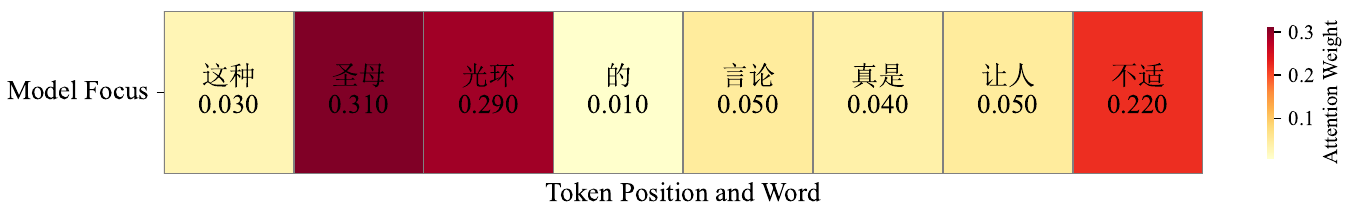}}
	\caption{Attention Weight Heatmap of Chinese Toxic Comments Based on the BiLSTM-Attention Model.}
	\label{fig4}
\end{figure}

To illustrate this method concretely, a simple experiment is designed in this section to simulate the entire process of attention visualization for Chinese toxic comments. The experiment is built on the PyTorch framework, where a BiLSTM model with an attention layer is constructed, and weight visualization and analysis are performed on a highly representative implicit toxic comment. As shown in Fig. \ref{fig4}, the visualization analysis of the comment "\begin{CJK}{UTF8}{gbsn}这种圣母光环的言论真是让人不适\end{CJK} [Such remarks with a holier-than-thou aura are really uncomfortable]" reveals that the model's attention is highly concentrated on the core metaphorical terms "\begin{CJK}{UTF8}{gbsn}圣母\end{CJK} [holier-than-thou]" and "\begin{CJK}{UTF8}{gbsn}光环\end{CJK} [aura]", assigns secondary weights to the negative term "\begin{CJK}{UTF8}{gbsn}不适\end{CJK} [uncomfortable]", and ignores grammatical words. This verifies that the trained model can identify implicit toxicity in Chinese online contexts, and its decision-making logic is highly consistent with human judgments.

This visualization method also possesses model diagnostic value. Abnormal attention distributions (e.g., the model over-focusing on place names or neutral words) can effectively reveal biases in training data and provide clear guidance for data cleaning. In practical applications, heatmaps can be integrated into moderation interfaces to support human-machine collaborative decision-making. Moderators can quickly review cases based on the rationality of attention focus, and edge cases can be fed back into the model optimization process, forming a sustainable closed loop of detection-interpretation-optimization.

\noindent\textit{(4) Attribution Quantification in Pre-trained Models}

Pre-trained models represented by BERT feature massive parameter scales and deeply complex architectures, making them typical black-box systems. Interpreting their decision-making processes requires more powerful attribution techniques. Integrated gradients \cite{IG} is a classic feature attribution method that assigns contribution scores of each input word to the final prediction by calculating the integral of gradients of input features (embedding vectors of each word) along the path from baseline values to actual values.

\begin{figure}[t]
	\centerline{\includegraphics[width=3.6in,keepaspectratio]{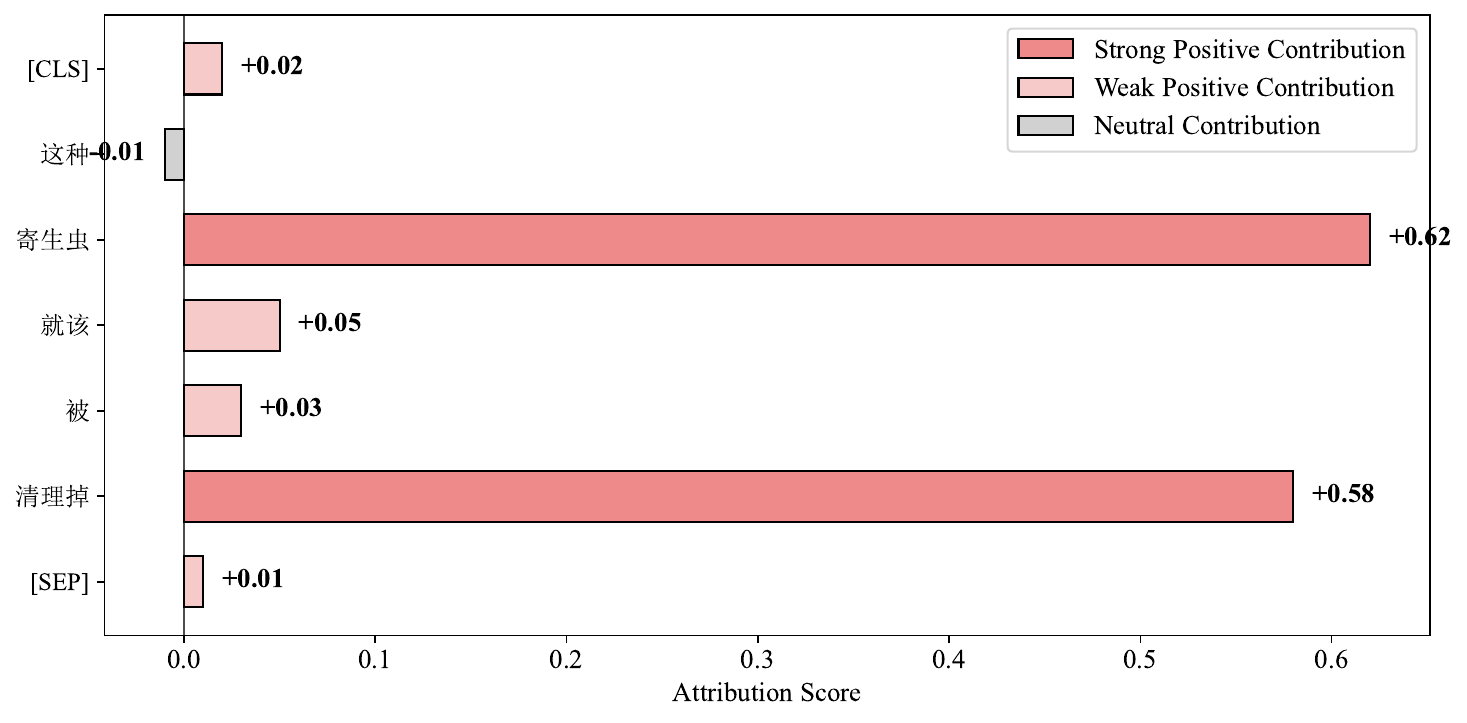}}
	\caption{Attribution Analysis of BERT Model Predictions Based on Integrated Gradients.}
	\label{fig5}
\end{figure}

Fig. \ref{fig5} presents a comparison of predicted attribution scores for the comment "\begin{CJK}{UTF8}{gbsn}这种寄生虫就该被清理掉\end{CJK} [Such parasites deserve to be eliminated]", analyzed by Integrated Gradients on a fine-tuned BERT model. The results show that the attribution scores of "\begin{CJK}{UTF8}{gbsn}寄生虫\end{CJK} [parasites]" and "\begin{CJK}{UTF8}{gbsn}清理掉\end{CJK} [eliminated]" are significantly the highest. These two terms are the core semantic units in the comment that carry the derogatory metaphor and violent tendency, and their combination constitutes the harmful implicit intent of the text; thus, the model allocates the main contribution scores to this phrase-level semantic unit. In contrast, words such as "\begin{CJK}{UTF8}{gbsn}这种\end{CJK} [this kind of]", "\begin{CJK}{UTF8}{gbsn}就该\end{CJK} [deserve to]" and "\begin{CJK}{UTF8}{gbsn}被\end{CJK}[be]" are demonstrative function words, modal auxiliary words or grammatical particles. They do not carry core semantic meanings and thus only obtain weak contribution scores. In addition, the attribution scores of [CLS] and [SEP] are extremely low because these two are structural tokens of BERT. [CLS] serves as a semantic aggregation carrier rather than an input semantic source, while [SEP] acts as a sequence separator rather than a semantic component. Neither of them carries the core intent of the comment, hence their negligible contributions. This visualization result clearly demonstrates that the decision-making of pre-trained models is not based on simple term frequency statistics; instead, it can capture the phrase-level deep semantics formed by word combinations, thereby identifying the implicit harmful intent of the text.

In the work of \cite{COLA}, interpretability tools such as integrated gradients are leveraged to directly quantify the contribution of input features to model classification results; combined with fine-grained annotated data, this approach renders the decision-making basis of the model traceable. Furthermore, the cutting-edge approaches to enhancing the interpretability of pre-trained models focus on fine-grained annotation and external knowledge empowerment. Representative works COLD \cite{COLD} and TOXICN \cite{TOXICN} provide differentiated practices: by constructing a dataset with multi-category fine-grained annotations, COLD offers category-level attribution references for model detection results, and indirectly supports interpretability analysis by identifying key factors (e.g., keywords related to specific groups) that trigger model judgments. TOXICN goes a step further: while building a dataset with a three-level annotation system of toxicity type-target group-expression category, it also provides a specialized lexicon of implicit insults, and explicitly integrates such external knowledge into model representations via a toxicity knowledge enhancement benchmark. This is equivalent to implanting explicit moderation rules into the model, thereby directly improving the transparency and interpretability of its decision-making process.

\noindent\textit{(5) LLMs: Chain-of-Thought Prompting and Generative Explanation Paradigms
}

The emergence of LLMs has spawned a novel generative explanation paradigm based on natural language interaction. Through well-designed prompt engineering, LLMs can be directly required to provide reasoning processes alongside their judgments. One of the most effective approaches is to stimulate their chain-of-thought capability. For instance, the following prompt can be provided to GPT-4: "Please analyze step by step whether the following comment is toxic and finally give a conclusion of 'Yes' or 'No'. Comment: \begin{CJK}{UTF8}{gbsn}您这大局观, 不愧是团队短板\end{CJK} [With such a holistic perspective, you truly live up to being the team's weakness]." The LLM may generate the following explanation:

\begin{tcolorbox}
\textbf{Step 1:} Understand the literal meaning. The comment praises the target for having a "\begin{CJK}{UTF8}{gbsn}大局观\end{CJK} [holistic perspective]", yet labels them as the "\begin{CJK}{UTF8}{gbsn}团队短板\end{CJK}  [the team's weakness]".

\textbf{Step 2:} Analyze contextual conflicts. In a team context, the term "\begin{CJK}{UTF8}{gbsn}短板\end{CJK} [weakness]" generally refers to someone who lacks competence and holds the team back, which contradicts the seemingly complimentary expression.

\textbf{Step 3:} Identify rhetorical devices. This comment most likely employs the rhetorical device of irony; its actual implication is to criticize the target for lacking a holistic perspective and impeding the team’s progress.

\textbf{Step 4:} Make a comprehensive judgment. By means of irony, the comment launches a personal attack and carries distinct negative sentiment and derogatory connotations.

\textbf{Conclusion:} Yes.
\end{tcolorbox}

However, the original chain-of-thought prompting still suffers from issues such as unstable reasoning processes and non-explicit justifications. To address this, cutting-edge research is committed to systematizing and explicating such inherent reasoning capabilities through structured prompting and external knowledge injection, so as to further improve the transparency and credibility of decision-making. 

For instance, the work \cite{CangjieToxi} improves interpretability via a multi-stage prompting framework, decouples character anomaly detection, semantic restoration and toxic content classification, clarifies the task objective of each step, and selects character replacement candidates by combining perturbation principles and contextual reasoning, thereby rendering the decision-making process of large models more transparent. The work \cite{ChineseHarm-Bench} enhances interpretability by constructing a manually annotated knowledge rule base, which defines the judgment criteria for six violation categories (including keywords and behavioral characteristics). It integrates this rule base and the implicit knowledge of the teacher model into prompt engineering and model training, thereby equipping large models with clear rule-based justifications for decision-making; meanwhile, the design of structured prompts further improves the transparency of the detection process.

\noindent\textit{(6) Conclusion} 

Based on the above review, we argue that in practice, a mature toxic comment detection system often adopts a strategy of multi-level interpretability collaboration. Firstly, it utilizes fast rules for preliminary filtering and interpretation; secondly, it employs DL or pre-trained models to handle complex cases, supplemented by visual attribution for moderators' review; finally, it submits the most challenging edge cases to LLMs to obtain in-depth generative explanations for final adjudication or integration into the knowledge base. This framework that organically integrates interpretability technologies at different levels is the key technical pathway for realizing the evolution of detection systems from static, closed black boxes to dynamic, open, and sustainably evolving human-machine collaborative white boxes.

\section{Open Problems and Challenges} \label{5}
Current research has achieved remarkable progress in Chinese online toxic comment detection, focusing on three dimensions: data engineering (e.g., constructing more balanced and fine-grained datasets), feature engineering (e.g., mining implicit expressions, Chinese character variants and prompt engineering), and model engineering (e.g., designing dedicated architectures to improve accuracy). However, these efforts generally share a common research limitation: most existing studies take isolated text segments divorced from their contextual presentation as the analysis objects.

\subsection{Homogeneous Context}
As a matter of fact, toxicity is not an abstract textual attribute but is highly dependent on its sociotechnical context. Beyond the fact that different platforms generate distinctly different toxic expressions due to their varied user groups, community norms and interaction modes, post topics also profoundly shape the emotional tendencies and attack boundaries of comments \cite{SCCD}. Current models lack the ability to model such macro-contexts, leading to a sharp decline in generalization performance when applied across platforms and topics.

\textit{Research Opportunities:}

\begin{itemize}
	\item Topic and Style Awareness: Investigate the automatic identification of comments' meta-information (e.g., publishing platform, topic section, and post theme), and integrate such information into models as conditional inputs or prior knowledge. For instance, develop adapter or prompt learning techniques to enable foundation models to dynamically adjust the judgment thresholds for linguistic styles in different contexts.
	\item Graph Neural Networks and Community Modeling: Treat comments not as isolated nodes but as components embedded in user-comment interaction graphs. Leverage graph neural networks to analyze users' historical behaviors, their positions in social networks, and comment propagation patterns, so as to identify organized aggressive behaviors or toxicity diffusion patterns influenced by community atmospheres.
\end{itemize}

\subsection{Single Modality}
Online communication is inherently multimodal, and single-modal text models cannot effectively address the increasingly complex and prevalent multimodal toxic attacks. Such attacks usually achieve malicious intentions through the interaction and combination of different modalities. For example, in image-text satirical scenarios, a superficially neutral comment like "Well done" paired with a highly offensive "Panda Head" meme has its toxic core entirely carried by the image, rendering text models completely ineffective in such cases. Ignoring multimodal signals will not only lead to the missed detection of a large number of such toxic comments, but more crucially, prevent models from understanding the composite semantics and deep-seated aggressive intentions jointly constructed by text, images, audio and other elements.

\textit{Research Opportunities:} Unlike the efforts on English toxic comment datasets \cite{multimodality}, the construction of high-quality Chinese multimodal datasets is a primary and urgent foundational task. This requires collecting and annotating large-scale, high-quality paired Chinese toxic comment data (text, images, videos), and designing a fine-grained annotation system—for example, how toxicity is manifested separately in text and images, as well as the overall aggressiveness after fusion. Subsequently, a unified feature fusion and decision-making framework should be explored to achieve joint reasoning in an end-to-end manner, enabling the judgment of the overall toxicity of multimodal content instead of simply performing post-hoc fusion on the results of each modality.

\subsection{Data Noise}
The performance of current Chinese toxic comment detection models is limited by inherent flaws in training data: the subjectivity of data annotation and the complexity of Chinese semantics jointly introduce systematic noise into datasets. All existing studies heavily rely on annotators' subjective perceptions to determine toxicity, and the strong context dependence, internet memes and implicit expressions of Chinese further amplify annotation inconsistency. This causes models to capture such annotation biases rather than genuine toxicity patterns during training, resulting in severe overfitting issues. The essence of the problem is that we use vague criteria fraught with human subjective divergences to train deterministic mathematical models, and deploy these models on Chinese—a complex, dynamic and context-dependent system. Inevitably, the upper limit of model performance is constrained by data quality.

\textit{Research Opportunities:} To address the above issues, hierarchical annotation guidelines integrating sociolinguistic and computational social science perspectives should be developed. These guidelines should not only label the presence of toxicity, but also annotate its target of attack, intent, expressive means and cultural sensitivity. Relative consensus can be established through multiple rounds of annotation, expert arbitration and dynamic calibration, which explicitly characterizes the distribution of subjectivity and provides richer learning signals for models. In addition, the reasoning capabilities of LLMs can be leveraged to identify, cluster and analyze inconsistently labeled samples in training data, assisting humans in data cleaning or reassignment of probabilistic labels.

\section{Conclusion}\label{6}
In this review, we conducted a comprehensive overview of NLP-based detection research on toxic comments in the Chinese cyberspace. The overview mainly covers the connotation and characteristics of Chinese toxic comments, with an analysis of the platform ecosystems and transmission mechanisms such comments rely on. Subsequently, existing research on Chinese toxic comment detection, including dataset construction and model design, was sorted out and critically analyzed. Notably, we proposed a novel fine-grained and scalable framework for toxic definition and classification, with elaboration on the efficient construction of datasets and the quality assessment after construction. In addition, the interpretability methods for NLP-based black-box detection models were summarized. Finally, an in-depth exploration was carried out on the three major challenges faced by current Chinese toxic comment detection work, together with the proposal of targeted research opportunities.

\bibliographystyle{ieeetr} 
\bibliography{MyRefs} 
~~~\\
~~~\\

\end{document}